\documentclass[]{tPHM2e}
\usepackage{amssymb}
\DeclareGraphicsExtensions{.pdf}
\begin{document}
\doi{10.1080/14786435.20xx.xxxxxx}
\issn{1478-6443}
\issnp{1478-6435}
\jvol{00} \jnum{00} \jyear{2010} 

\markboth{Taylor \& Francis and I.T. Consultant}{Philosophical Magazine}


\title{The Ising version of the $t$--$J$ model}

\author{Maciej M. Ma\'ska$^{{\rm a}\ast}$,\thanks{$^\ast$Email: maciej.maska@us.edu.pl}
Marcin Mierzejewski$^{\rm a}$ and E. Kochetov$^{\rm b}$\\
\vspace{6pt} $^a$\em{Department of Theoretical Physics, Institute of Physics, University of Silesia, 40--881 Katowice, Poland}; $^b$Theoretical Physics Laboratory, Joint Institute for
Nuclear Research, 141980 Dubna, Russia
\\\vspace{6pt}\received{v4.5 released May 2010} 
}

\maketitle

\begin{abstract}
The $t$--$J$ model is analysed in the limit of strong anisotropy, where the transverse components of electron spin are neglected. We propose a slave--particle--type approach that is valid, in contradiction to many of the standard approaches, in the low--doping regime and becomes exact for a half--filled system. We describe an effective method that allows to numerically study the system with the no--double--occupancy constraint rigorously taken into account at each lattice site. Then, we use this approach to demonstrate the destruction of the antiferromagnetic order by increasing doping and formation of Nagaoka polarons in the strong interaction regime.
\bigskip
\begin{keywords}$t$--$J$ model; antiferromagnetism; Nagaoka polaron; strong correlations. 
\end{keywords}\bigskip
\end{abstract}

\section{Introduction}
It is commonly believed that the richness of the behaviour of strongly correlated systems is a result of a competition between the kinetic and interaction energies \cite{anderson1959}. 
Unfortunately, due to the presence of strong correlations many of the ''traditional'' solid state methods, like the density functional theory within local density approximation, or many-body perturbation theory, that handled impressively well simple metals, covalent semiconductors, closed-shell ionic insulators, and even intermetallic compounds, cannot be used. It has been recognized for many years that strongly--correlated systems require a distinct paradigm from what was successful for the mentioned above systems.

It is also believed that the essence of the physics of the strongly correlated systems can be described by simple one--band Hamiltonians that are able to properly take into account the competition between the kinetic and interaction energies. Two of the most acceptable models are the Hubbard model \cite{hubbard} and its effective strong--interaction version, namely the $t$--$J$ model \cite{spalek1977,spalek1978}. Both these models contributed greatly to our understanding of strongly correlated systems. Unfortunately, apart from some specific cases, none of these models can be solved exactly. Therefore, it is very important to develop analytical or numerical methods that can be applied to systems described by interacting Hamiltonians. Moreover, it is equally important to be able to determine the errors introduced by the applied approximation.

\subsection{The Hubbard and $t$--$J$ models}
The Hamiltonian of the Hubbard model, originally introduced to describe correlation effects in narrow $d$--band materials, has the following form:
\begin{equation}
H_{\rm Hubb}=-t\sum_{{\langle ij\rangle}\sigma}c^\dagger_{i\sigma}c_{j\sigma}+U\sum_in_{i\uparrow}n_{i\downarrow},
\label{hubbard}
\end{equation}
where the first term describes the kinetic energy and the second the interactions. Here, $c^\dagger_{i\sigma}$ ($c_{i\sigma}$) creates (annihilates) an electron of spin $\sigma$ at site $i$ and the occupation number operator $n_{i\sigma}\equiv c^\dagger_{i\sigma}c_{i\sigma}$. The Hilbert space of the Hubbard model contains four states per site: $|\Theta\rangle$, $|\uparrow\rangle$, $|\downarrow\rangle$ and $|\uparrow\downarrow\rangle$.

Since in the Hubbard model there is only an on--site interaction, in the limit of large $U$ it is energetically very expensive for electrons to hop onto already occupied sites. Therefore, for the average occupation less or equal to one electron per lattice site the low energy processes take place mainly in the lower Hubbard subband. However, virtual excitations with double occupied sites may increase the electron mobility leading to lowering the total energy. The effective Hamiltonian can be derived from the strong coupling expansion of the Hubbard model with respect to $t/U$. It was shown that that virtual excitations generates a spin--spin exchange interaction between neighbouring sites, the so--called kinetic exchange. After the transformation the states in the lower Hubbard band are described by the $t$--$J$ model acting in a projected Hilbert space containing only three states per site: $|\Theta\rangle$, $|\uparrow\rangle$, $|\downarrow\rangle$. The state $|\uparrow\downarrow\rangle$ is removed by the Gutzwiller projection operator. The Hamiltonian of the $t$--$J$ model is given by:
\begin{equation}
H_{t-J}=-t\sum_{\langle ij\rangle\sigma} \tilde{c}_{i\sigma}^{\dagger}
\tilde{c}_{j\sigma}+ J\sum_{\langle ij\rangle} \left(\bm S_i \bm S_j -
\frac{1}{4}\tilde{n}_i\tilde{n}_j\right),\label{tJ}
\end{equation}
where the antiferromagnetic exchange constant $J=4t^2/U$. $\tilde{c}_{i\sigma}$ ($\tilde{c}_{i\sigma}^{\dagger}$) represents fermionic annihilation (creation) operators projected onto a space without double occupancy: $\tilde{c}_{i\sigma}=(1-n_{i,-\sigma})c_{i\sigma}$. Despite a potential inadequacy of this model to represent real strongly correlated materials, it is still the simplest model that captures the important antiferromagnetic correlations of weakly doped antiferromagnets. Thus, it is crucial that the properties of this model are well understood. The Hamiltonian given by Eq. (\ref{tJ}) has been investigated intensively by different analytical and numerical methods. The analytical methods are usually limited to only one or two holes in an antiferromagnetic background. It is very difficult to treat in a systematic non--perturbative way systems with strong correlations. In the case of the $t$--$J$ model an additional difficulty comes from the fact that the operators $\tilde{c}_{i\sigma}$ and $\tilde{c}_{i\sigma}^{\dagger}$ do not fulfil the usual fermionic commutation rules. This non--fermionic behaviour results, in turn, from the projection of the states with doubly occupied lattice sites. Unfortunately, the constraint of no double occupancy becomes very important close to half filling and only methods which are capable of taking it into account without uncontrollable approximations can give reliable results in this regime. And this is a regime of particular interest because the high--temperature superconductors are slightly doped antiferromagnets. 

Due to the difficulties in analytical approaches, numerical methods such as exact diagonalization, density-matrix renormalization group and quantum Monte-Carlo are extensively performed to study this model. The exact diagonalization can only be performed in a very small lattice size, and the density-matrix renormalization group method is largely restricted to one--dimensional systems. In contrast, Quantum Monte Carlo simulation is the only systematic and scalable method with sufficient numerical accuracy for higher dimensional problems. However, this method also has the notorious fermion sign problem which makes low temperature properties inaccessible.

\subsection{Slave--particle approaches to the $t$--$J$ model}
The single occupancy constraint, that makes analytical approaches to the $t$--$J$ model so difficult can be written as
\begin{equation}
\sum_{\sigma}c^\dagger_{i\sigma}c_{i\sigma}\le 1,
\label{constr}
\end{equation}
for every lattice site $i$. In order to treat this constraint in a controllable way a number of slave--particle methods have been proposed \cite{Baskaran1987973,PhysRevLett.58.2790,Kotliar88,JPSJ.57.2768,RevModPhys.78.17}. In the slave particle formalism, the electron operator is expressed in terms of auxiliary fermions and bosons. For instance, in the slave boson formalism the electron annihilation operator
$c_{i\sigma}$ is given by 
$c_{i\sigma}=b^{\dagger}_{i}f_{i\sigma}$,
where $b^{\dagger}_{i}$ is a boson operator and $f_{i\sigma}$ is a fermion operator. In the slave fermion representation $c_{i\sigma}=b^{\dagger}_{i\sigma}f_{i}$. Instead of the difficult to handle constraint of Eg. (\ref{constr}), one considers more convenient slave--particle constraints 
\begin{equation}
b^{\dagger}_ib_i+\sum_{\sigma}f^{\dagger}_{i\sigma}f_{i\sigma}=1\ \ \ \mbox{or}\ \ \ \
\sum_{\sigma}b^{\dagger}_{i\sigma}b_{i\sigma}+f^{\dagger}_{i}f_{i}=1,
\end{equation}
where the fermion (boson) operator keeps track of the spin and the boson (fermion) operator keeps track of the charge in the case of the slave--boson (slave--fermion) representation. Such slave--particle approaches are usually studied in a functional integral representation of the partition function with the no double occupancy constraints enforced with the help of Lagrange multiplier. To solve the problem the mean field approximation is usually applied
and the Lagrange multiplier is taken to be independent of the lattice site. It means, however, that the local no double occupancy constraint is replaced by a global one with uncontrollable consequences. 

\section{The Ising version of the $t$--$J$ model}
Because of the difficulties in solving the full $t$--$J$ model, often its simplified versions are studied. One of them is the $t$--$J_z$. This model can be considered as a limiting $(J_{\bot}=0)$ case of the $t$--$J$ model (\ref{tJ}) which has an Ising rather than a Heisenberg spin interaction:
\begin{equation}
H_{t-J_z}=-t \sum_{\langle ij\rangle\sigma} \tilde{c}_{i\sigma}^{\dagger}
\tilde{c}_{j\sigma}+ J_z\sum_{\langle ij\rangle} \left( S^z_i S^z_j -
\frac{1}{4}\tilde{n}_i\tilde{n}_j\right).
\label{tJz}
\end{equation}
The original $t$--$J$ model possesses the continuoues global SU(2) spin symmetry. In the Hamiltonian (\ref{tJz}) the interaction term $S^z_i S^z_j$ has a lower discrete Z$_2$ symmetry. The rest of the terms, however, still possess the original SU(2) symmetry. As a result, the total symmetry of the $t$--$J_z$ Hamilonian is dependend on the value of the $J_z$ coupling. For $J_z=0$ the symmetry is SU(2) like in the full $t$--$J$ Hamiltonian, whereas for $J_z=0$ it is only Z$_2$.

In contradistiction to the $t$--$J_z$ Hamilonian, all terms of the full Ising--$t$--$J$ Hamiltonian possess only the discrete $Z_2$ symmetry, independetly of the values of the model parameters. Unfortunately, since the operators $\tilde{c}_{i\sigma}$ transform themselves in the fundamental representation of SU(2), there is no obvious way to derive it from the $t$--$J$ model (\ref{tJ}). One of the possibility is to use the enlarged spin-dopon representation of the operators $\tilde{c}_{i\sigma}$ \cite{mmfk2009}.

In the framework of this approach fermion operators $d_{i\sigma}$ are assigned to doped carriers (like holes) instead to the lattice electrons. The vectors that span the enlarged on--site Hulbert space have the form of $|\sigma a\rangle$, where  $\sigma=\Uparrow,\Downarrow$ labels the 2D lattice spin $\bm{Q}_i$ Hilbert space and $a=0,\ \uparrow,\ \downarrow,\ \uparrow\downarrow$ labells the 4D onsite dopon Hilbert space. The physical subspace is spanned by the spin-up $|\Uparrow 0\rangle_i$, spin-down $|\Downarrow 0\rangle_i$, and spinless vacancy $\left(|\Uparrow \downarrow\rangle_i - |\Downarrow \uparrow\rangle_i\right)/\sqrt{2}$ states \cite{ribeiro2005}. 
\begin{figure}
\begin{center}
\includegraphics[width=0.45\linewidth,,natwidth=2030.77,natheight=876.8]{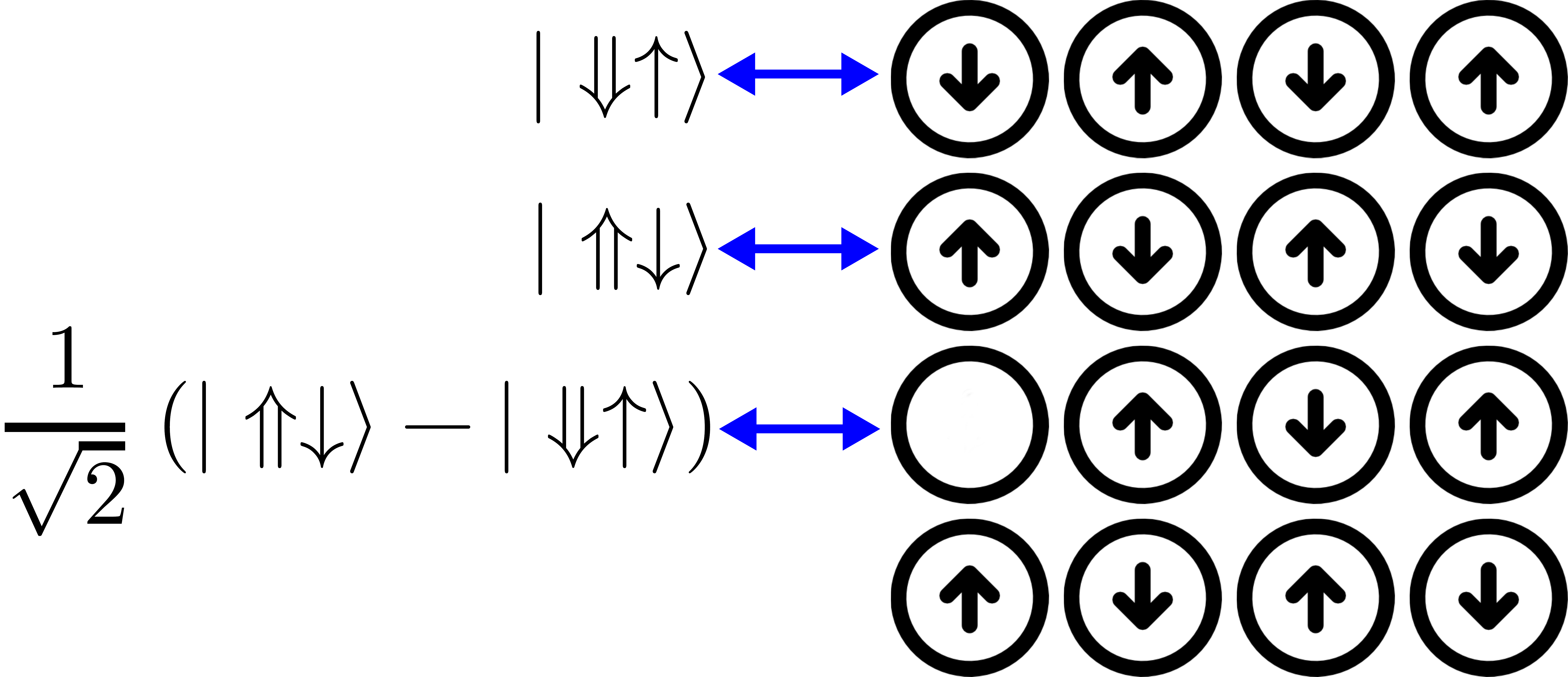}
\caption{Mapping between the physical states from the Hilbert space with no double occupied 
sites and the lattice spin and dopon states.}%
\label{representat}
\end{center}
\end{figure}
The constraint
\begin{eqnarray}
\bm{Q}_i\bm{M}_i +\frac{3}{4}n^d_i =0 \label{2.3},
\end{eqnarray}
has to be applied to remove the remaining unphysical states \cite{fku2007}.
In the above $\bm M_i=\sum_{\sigma,\sigma'}{d}_{i\sigma}^{\dagger}
\bm\tau_{\sigma\sigma'}{d}_{i\sigma'}$ is the dopon spin operator. This way the physical spin operator can be expressed as
\begin{equation}
\bm{S}_i=\bm{Q}_i+\bm{M}_i.
\label{Q}\end{equation}
Taking into accont that $(Q^{\alpha})^2=\frac{1}{4}$, Eq. (\ref{2.3}) can be written as
\begin{eqnarray}
\sum_{\alpha=x,y,z} Q^{\alpha}_i{M}^{\alpha}_i +n^d_i\!\!\!\sum_{\alpha=x,y,z}(Q^{\alpha}_i)^2 =0
\label{2.3*}.
\end{eqnarray}
In the full Ising $t$--$J$ model the transverse spin components should vanish identically.
According to Eq. (\ref{Q}), this requires $Q_i^{\pm}=M_i^{\pm}=0$. 
Then, the Ising $t$--$J$ model can be derived by projecting the dopon operators onto 
the Hilbert space determined by the local constraint
\begin{equation}
Q^z_iM^z_i+ \frac{1}{4}n^d_i=0.
\label{2.4}\end{equation}
that is the Ising counterpart of Eq. (\ref{2.3}).
The projected physical electron operators $\tilde{c}_{i\sigma}$ can be then expressed in terms of the lattice spin and dopon operators:
\begin{subequations}
\begin{eqnarray}
\tilde{c}_{i\downarrow}&=&{\cal P}^{\rm ph}_i d_{i\uparrow}^{\dagger}{\cal P}^{\rm ph}_i
=\left(\frac{1}{2}-Q^z_i\right) d_{i\uparrow}^{\dagger},\label{2.5} \\
\tilde{c}_{i \uparrow} &=& {\cal P}^{\rm ph}_i d_{i \downarrow}^{\dagger}{\cal P}^{\rm ph}_i=\left(\frac{1}{2}+Q^z_i\right)d_{i\downarrow}^{\dagger}
\label{2.5*}
\end{eqnarray}
\end{subequations}
where the operator ${\cal P}^{\rm ph}$ that projects onto the physical subspace is given by 
${\cal P}^{\rm ph}_i=1-(2Q^z_iM^z_i+\frac{1}{2}n^d_i)$.
Then, it can be easily shown that the Ising version of Eq. (\ref{Q}) is fulfilled:
\begin{equation}
 S_i^z=\frac{1}{2}(\tilde {c_{i \uparrow}}^{\dagger}\tilde {c_{i \uparrow}}
-\tilde c_{i\downarrow}^{\dagger}\tilde c_{i\downarrow})
=Q_i^z+M_i^z
\end{equation}
and the transverse components of the physical spin operators vanish identically.
\begin{equation}
S^{+}_i=(S^{-}_i)^{\dagger} =\tilde c^{\dagger}_{i\uparrow}\tilde c_{i\downarrow}\equiv 0.
\label{trans_spin}
\end{equation}

%

The adventage of the Ising representation of the $t$--$J$ model is
particularly visible close to half--filling, where the Hamiltonian (\ref{tJz})
is reduced to the following form \cite{mmfk2009}:
\begin{equation}
H^{\rm Ising}_{t-J} = t\sum_{\langle ij\rangle\sigma}
d_{i\sigma}^{\dagger} d_{j\sigma}
+  J\sum_{\langle ij\rangle}\left[\left(Q^z_i Q^z_j-\frac{1}{4}\right)
 +  Q_i^zM^z_j +Q_j^zM^z_i \right],
\label{2.6}
\end{equation}
which has to be accompanied by Eq. (\ref{2.4}).
Since close to half--filling the hole concentration $\delta$ is small, in Eq. (\ref{2.6}) 
we could drop the term describing the direct inter--dopon spin--spin interaction 
$M_i^zM^z_j$, which is proportional to $\delta^2$.

Since $[Q^z_i,\:H^{\rm Ising}_{t-J}]=0$ 
the spin degrees of freedom in Eq. (\ref{2.6}) can be described by classical variables. 

The constraint (\ref{2.4}) can be enforced with the help of a Lagrange multiplier.
Since for each lattice site $i$ $Q^z_iM^z_i+ \frac{1}{4}n^d_i\ge 0$, 
the global Lagrange multiplier 
\begin{equation}
\lambda\sum_i\left(Q^z_iM^z_i+ \frac{1}{4}n^d_i\right)
\label{cons}
\end{equation}
ensures that the constraint (\ref{2.4}) is fulfilled locally and the occupancy of
an unphysical state at arbitrary site would lead to an increase of the total energy by 
$\lambda\to +\infty$.
As a result, all unphysical states are eliminated, so that the constraint (\ref{2.4})
fulfilled rigorously.

The Hamiltonian (\ref{2.6}) accompanied by the constraint (\ref{cons}) represents a system described by 
classical ($Q^z_i$) as well as quantum ($d_i$) degrees of freedom. 
However, as pointed out above, the direct interaction between the quantum particles can be neglected close
to half--filling and only the interaction between quantum and classical particles is present in Eq. (\ref{2.6}).
In this aspect, the Ising $t$--$J$ model is similar to the Falicov--Kimball model and efficent hybrid methods
that have been developed for latter model can be applied.

\section{Numerical approach}
The numerical technique we use to solve the effective model is based on a method that combines Monte Carlo simulations with exact diagonalization of one--particle Hamiltonians. This technique was proven to work effectively for the Falicov--Kimball model \cite{mm1,mm2,mm3}. The details of the application of this method to the Ising $t$--$J$ model are described in Ref. \cite{mmfk2009}, here we will sketch it for the sake of completeness. 

The Hamiltonian of the Ising $t$--$J$ model given by Eq. (\ref{2.6}) can be divided into a one--particle part describing itinerant quantum particles with atomic levels varying from site to site and a part describing Ising--type interactions between the classical variables $Q^z_i$. The values of the atomic levels is determined by the distribution of the variables $Q^z_i$. Together with the Lagrange multiplier term the Hamiltonian can be written as
\begin{equation}
H^{\rm Ising}_{t-J}(\lambda)=\sum_{ij\sigma}{\cal T}_{ij\sigma}(\lambda)d^{\dagger}_{i\sigma}d_{j\sigma}+J\sum_{\langle ij\rangle}Q^z_iQ^z_j+{\rm const}.
\label{tJ_Is}
\end{equation}
The hopping matrix ${\cal T}_{ij\sigma}(\lambda)$ is given by
\begin{equation}
{\cal T}_{ij\sigma}(\lambda)=t_{ij}+\delta_{ij}\left\{\lambda\left[\frac{1}{2}+s(\sigma)Q^z_i\right]+s(\sigma)\frac{J}{2}\sum_{\langle j\rangle_i}Q^z_j\right\}.
\label{hop_mat}
\end{equation}
Since close to half filling details of the dispersion relation are important in strongly correlated systems \cite{PhysRevB.52.4597,PhysRevB.54.R12653,PhysRevB.54.10125,PhysRevLett.90.067001}, in the above equation we used $t_{ij}$ instead of the nearest--neighbour hopping $t$. By choosing proper values of $t_{ij}\equiv t({\bm r}_i-{\bm r}_j)$ one can reproduce the dispersion relation of, e.g., high--$T_c$ superconductors. In numerical calculations we restrict 
the hopping range to third nearest neighbours, i.e., only $t,\ t'$ and $t''$ are nonzero.
$s(\sigma)$ is equal to 1 for $\sigma=\Uparrow$ and -1 for $\sigma=\Downarrow$;
$\langle j\rangle_i$ indicates that in the summation $j$ runs over all nearest neighbours of site $i$. Note, that the quantum and classical degrees of freedom are coupled by the exchange constant $J$ and by the Lagrange multiplier $\lambda$. Numerical simulations indicate that both these couplings may be important, e.g, $\lambda$ is crucial in destroying an antiferromagnetic order when the concentration of holes increases, whereas $J$ plays important role in formation of spin polarons.

For a given distribution $\{Q^z_i\}$ of the classical variables the hopping matrix (\ref{hop_mat}) can be numerically diagonalized and the Hamiltonian (\ref{tJ_Is}) can be rewritten as
\begin{equation}
H^{\rm Ising}_{t-J}(\lambda)=\sum_{n\sigma}{\cal E}_{n\sigma}\left(\{Q^z_i\},\lambda\right)d^{\dagger}_{n\sigma}d_{n\sigma}+J\sum_{\langle ij\rangle}Q^z_iQ^z_j,
\end{equation}
where the constant term was neglected. This form of the Hamiltonian allows to carry out the classical Monte Carlo simulations based on a modified Metropolis algorithm \cite{mm1,metropolis}. In the first step we choose an initial configuration $\{Q^z_i\}$. It is defined by the distribution the lattice spins with three possibilities at each site: spin up, spin down, empty. The number of empty sites is given by the doping level. Depending on the physical problem some additional constraints may be imposed on the initial state. For example, we may require equal numbers of spin--up and spin--down sites to run a simulation in a subspace of the total magnetization equal to zero $\sum_i Q^z_i=\sum_i M^z=0$. Next, the Hamiltonian (\ref{tJ_Is}) is diagonalized and the free energy of the dopons in the initial state is calculated. Then, we attempt to change the configuration $\{Q^z_i\}\:\rightarrow\: \{Q'^z_i\}$. The changes can be twofold: one can modify the direction of one or two lattice spins or the distribution of the empty sites can be altered. The decision what kind of attempt is made is random. In the case of spin modifications if we work in a subspace of zero total magnetization, we randomly choose two lattice sites with opposite spin directions and exchange the spins. Otherwise we simply flip a randomly chosen spin. After the modification of the state is made, the Hamiltonian (\ref{tJ_Is}) is again diagonalized, what gives the energy spectrum and the eigenstates of dopons. A new value of the dopon free energy is calculated and the configuration $\{Q'^z_i\}$ is accepted or rejected according to the Metropolis criterion. This criterion is modified with respect to the original that is used in simulations of classical systems: the internal energy in statistical weights is replaced by the free energy of the quantum subsystem (dopons). A detailed description of this approach can be found in Ref. \cite{mm1}. The Monte Carlo simulation gives all the characteristics of both the classical (lattice spins) and quantum (dopons) subsystems, i.e., all correlation functions, specific heat, magnetization, spectral functions, etc. can be determined as a function of temperature, doping level, interaction strength, dispersion relation, etc. Moreover, since we work in the real space we can study inhomogeneous systems, e.g., with polarons. Since in this approach only a one--particle Hamiltonian has to be diagonalized there is no limit to the size of the system from the available computer memory. The only limit comes from the CPU time, because in each Monte Carlo step the matrix given by Eq. (\ref{hop_mat}) is diagonalized, what significantly slows down the simulation in comparison to simulations of classical systems. Nevertheless, we are able to run simulations for 50$\times$50 lattices, what much beyond the capabilities of the fully quantum mechanical methods like the exact diagonalization based on the Lancz\"os algorithm or the Quantum Monte Carlo.

The simulations have been carried out in the canonical ensemble, which allows for accurate control of the concentration of holes. The unphysical states have been removed by the term (\ref{cons}) with $\lambda$ of the order of a few hundreds. This way $\lambda$ is by far the largest energy scale in the system, which guaranties the single occupancy of each lattice site.

One of the areas where the Ising $t$--$J$ model can be applied is the problem of the rapid suppression of the antiferromagnetic order with increasing doping level in the high--$T_c$ superconductors. In order to study the antiferromagnetic order we have to be able to calculate the spin--spin correlation function. In the Ising $t$--$J$ model it can be defined as
\begin{equation}
g(r)=\frac{4}{N^2}\sum_i\sum_j 
e^{i {\bm K}\cdot ({\bm R}_i-{\bm R}_j)}
\langle (Q^z_i+M^z_i) (Q^z_j+M^z_j) \rangle
\bar{\delta}(r-|{\bm R}_i-{\bm R}_j|),
\label{corfun}
\end{equation}
where ${\bm K}=(\pi,\pi)$ and
$$
\bar{\delta}(x)=\left\{\begin{array}{ll}
1 & {\rm if}\ |x|\le 0.5a, \\
0 & {\rm otherwise},
\end{array}\right.
$$
with $a$ being the lattice constant.
$\langle \ldots \rangle $ in Eq. (\ref{corfun}) means an average over the spin configurations generated in the Monte Carlo run. This quantity will allow to describe the character of the antiferromagnetic correlations. For a long--range order it will has a finite value for arbitrary distance $r$, for a quasi--long--range order it will decay algebraically, and for a short-range order it will decay exponentially. Another quantity which is easy to calculate is the static spin--structure factor, given by
\begin{equation}
S(\bm{q})=\frac{1}{N^2}\sum_{ij}e^{i \bm{q} \left(\bm{R}_i-\bm{R}_j \right)}
\langle (Q^z_i+M^z_i) (Q^z_j+M^z_j) \rangle .
\label{sq}
\end{equation}
What is more interesting, this modified classical Monte Carlo approach can give also dynamic properties of the dopons, which are fully quantum mechanical particles. Namely, one can calculate the dopon's spectral function 
\begin{equation}
A({\bm k},\omega) = -\frac{1}{\pi}{\rm Im}\: G \left({\bm k},\omega+i0^+\right),
\label{adef}
\end{equation}
where
\begin{equation}
G \left({\bm k},z\right) = \sum_{ij\sigma} e^{i{\bm k} \left( {\bm R}_i
- {\bm R}_j\right)}
\left\langle
{\cal G}_{\sigma}\left({\bm R}_i, {\bm R}_j,z \right)
\left[\frac{1}{2} -s(\sigma) Q^z_i\right]
\left[\frac{1}{2} - s(\sigma) Q^z_j\right]
\right\rangle,
\label{gdef}
\end{equation}
Here, similarly to Eqs. (\ref{corfun}) and (\ref{sq}), $\langle \ldots \rangle$ indicates
 averaging over spin configurations generated in the Monte Carlo runs and
\begin{equation}
{\cal G}_{\sigma}\left({\bm R}_i, {\bm R}_j,z \right)=
\left\{\left[z-\sum_{kl}{\cal T}_{kl\sigma}(\lambda)d^{\dagger}_{k\sigma}d_{l\sigma}
\right]^{-1}\right\}_{ij}
\end{equation}
is the real--space Green function for a given spin configuration $\{Q^z_i\}$. ${\cal T}_{kl\sigma}(\lambda)$ is given by Eq. (\ref{hop_mat}). Note that all the quantities given by Eqs. (\ref{corfun}), (\ref{sq}) and (\ref{adef}) are defined for physical electrons.

\section{Antiferromagnetism in the Ising $t$--$J$ model}
The evolution of the antiferromagnetic Mott insulating state into a superconducting state is one of the most intriguing problems in the physics of the high--$T_c$ superconductors. In particular, it is difficult to explain how the antiferromagnetic order is destroyed very quickly when charge carriers are doped into a parent cuprate material. In most thermodynamic measurements for hole doped cuprates, long range antiferromagnetism does not coexist with superconductivity and disappears completely around doping density $\delta \simeq 5\%$. Most of analytical and numerical studies of the $t$--$J$ model show that while upon doping the antiferromagnetic order is suppressed, it survives to much larger hole density that observed in experiments. The discrepancies may imply that the $t$--$J$ model is insufficient to describe the physics of the hight--$T_c$ superconductors. But they may also imply that the methods used to study this model close to half filling are not reliable enough to give the correct value of the critical hole density. It was already mentioned in the Introduction that most of both analytical and numerical methods have difficulties in dealing with the $t$--$J$ model close to half filling, where the constraint of no double occupancy is particularly important. This is the regime where we believe the validity of the Ising version of the $t$--$J$ model is most justified.

Most of the results for the $t$--$J$ model close to half filling are restricted to one or two holes in an antiferromagnetic background \cite{PhysRevLett.60.2793,PhysRevLett.60.740,
PhysRevB.37.1597,PhysRevB.58.15160,PhysRevB.41.9049,PhysRevB.42.10706,PhysRevB.44.317,
PhysRevB.47.14267,PhysRevB.52.R15711,PhysRevB.62.15480}. These methods do take into account the strong electron correlations, however, they do not allow to change the hole concentration and study the evolution of the antiferromagnetic order. On the other hand, variations of mean--field--type approaches \cite{PhysRevB.40.2610,0953-8984-6-27-022,
PhysRevB.83.104512,PhysRevB.79.214519,PhysRevB.81.073108,0295-5075-52-1-087,PhysRevB.59.8943,
PhysRevB.44.12077,PhysRevB.41.2653,PhysRevLett.80.2393} allow to control the hole density, but their validity is questionable in the underdoped regime, where electronic correlations are crucial due to the proximity to the Mott state. 

The proposed numerical approach to the Ising $t$--$J$ model takes advantages from both these groups of methods: on the one hand the concentration of holes can be changed almost continuously from zero to an arbitrary density, on the other hand the no--double--occupancy 
constraint is fulfilled rigorously not on average, like in the mean--field approaches, but at every lattice site. This was possible at the expense of neglecting the transverse spin--flip term (\ref{trans_spin}). However, it was shown in Refs. \cite{mmfk2009} and 
\cite{PhysRevB.85.245113} that the energy of one and two holes calculated for the full
$t$--$J$ model \cite{PhysRevB.64.024411,PhysRevB.76.035121,PhysRevLett.103.186401} and for its Ising version are close.

\begin{figure}
\begin{center}
\includegraphics[width=0.5\linewidth,natwidth=1030,natheight=773]{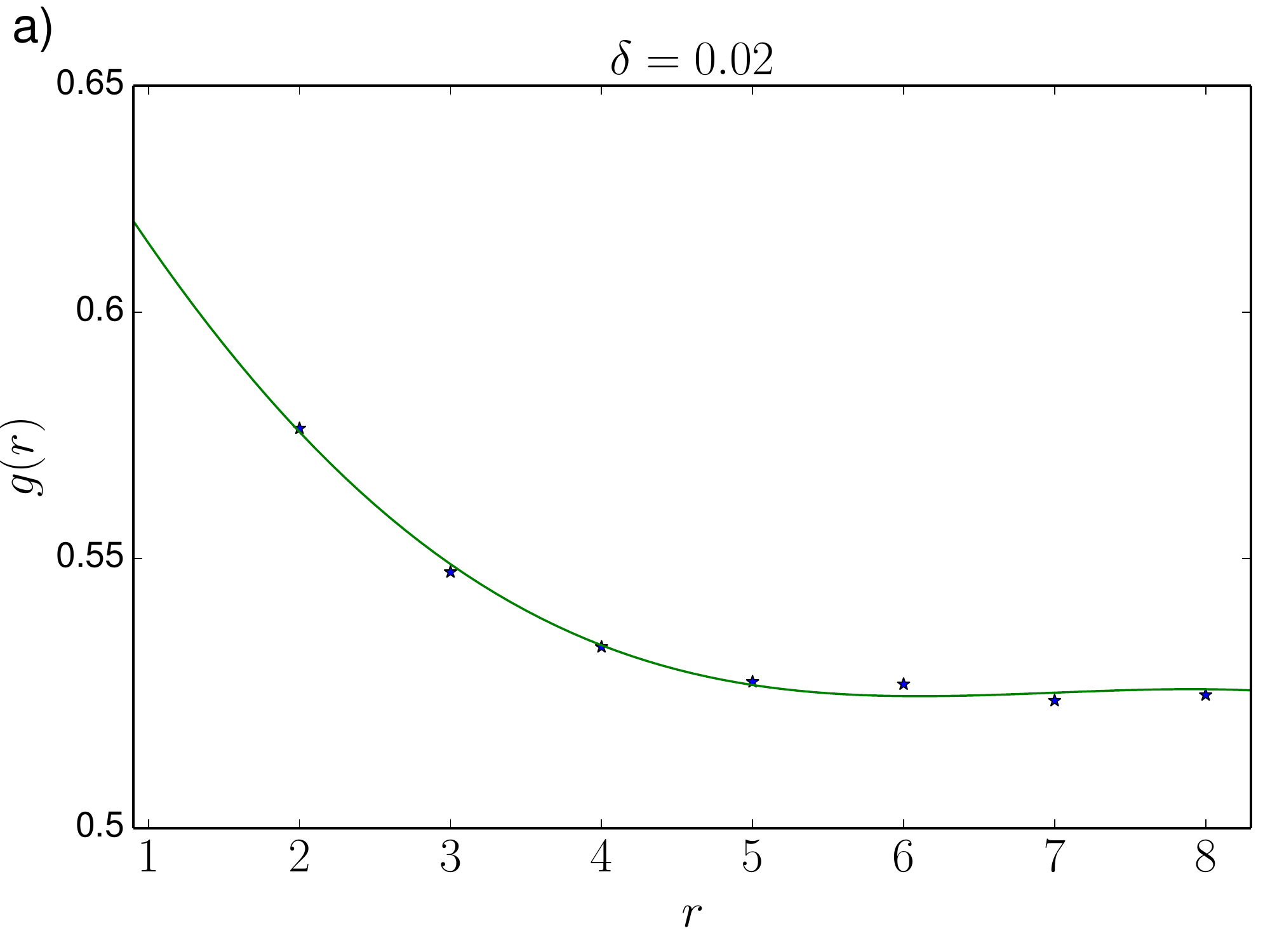}\includegraphics[width=0.5\linewidth,natwidth=1030,natheight=773]{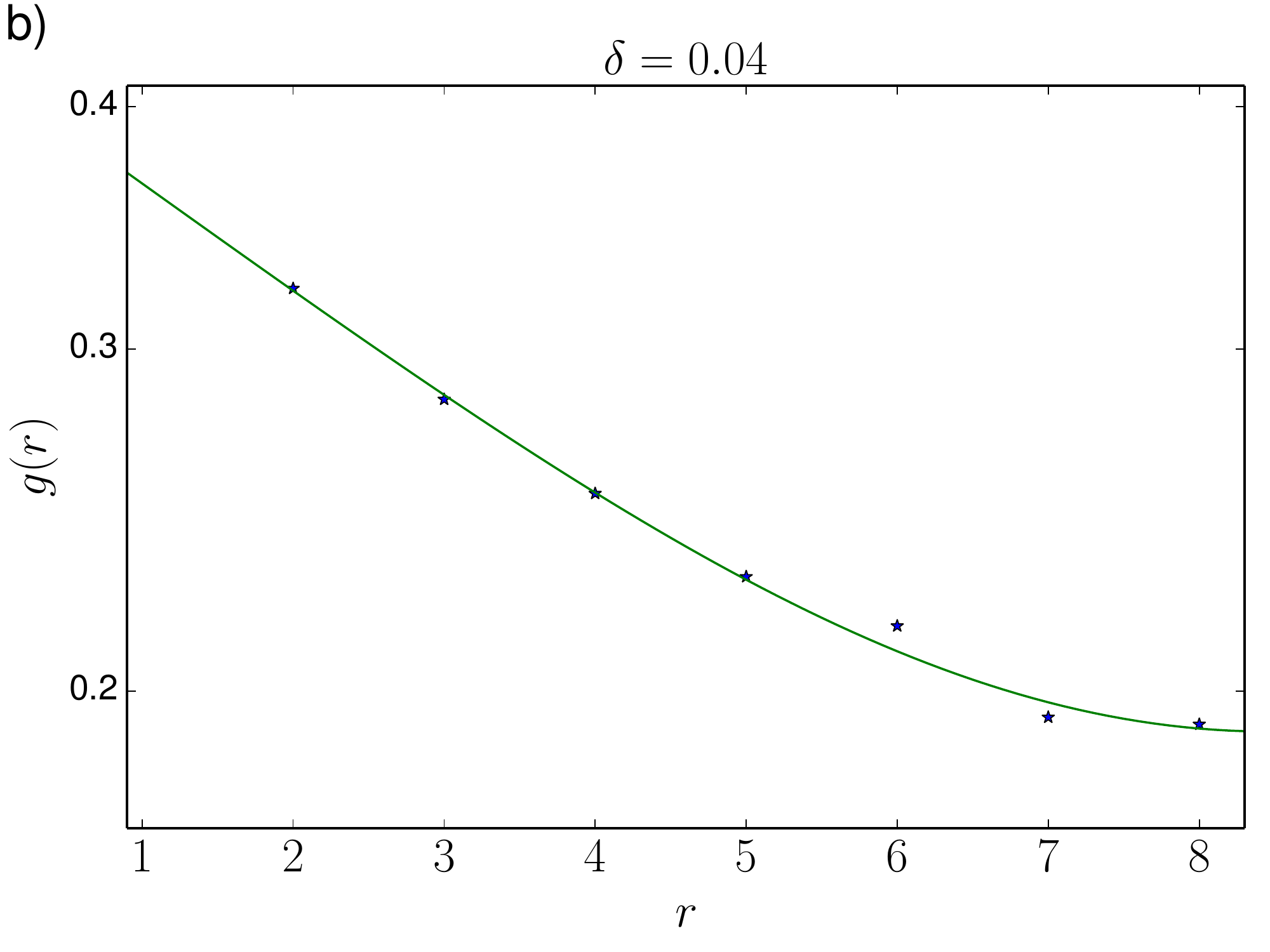}
\includegraphics[width=0.5\linewidth,natwidth=1030,natheight=773]{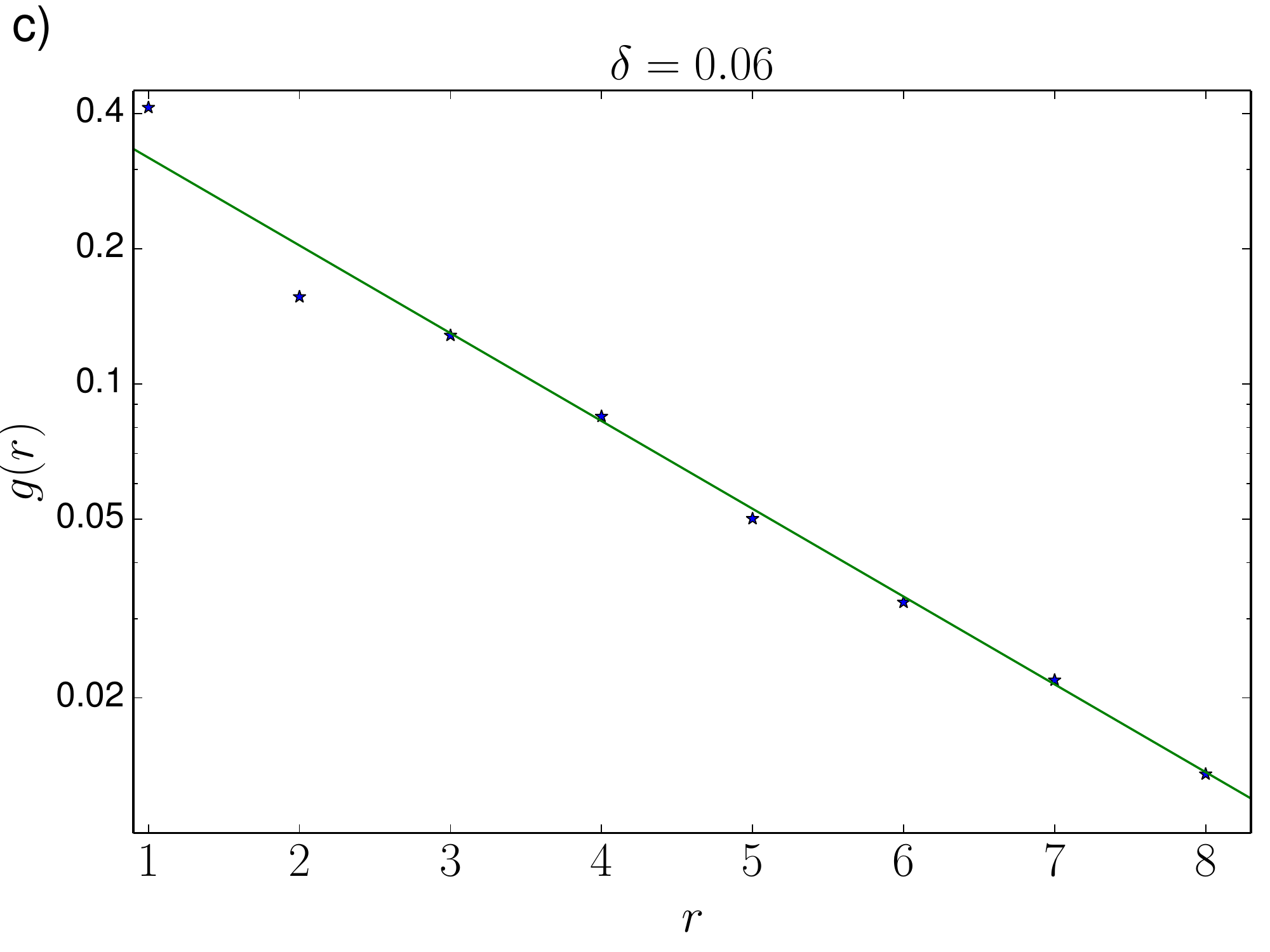}\includegraphics[width=0.5\linewidth,natwidth=1030,natheight=773]{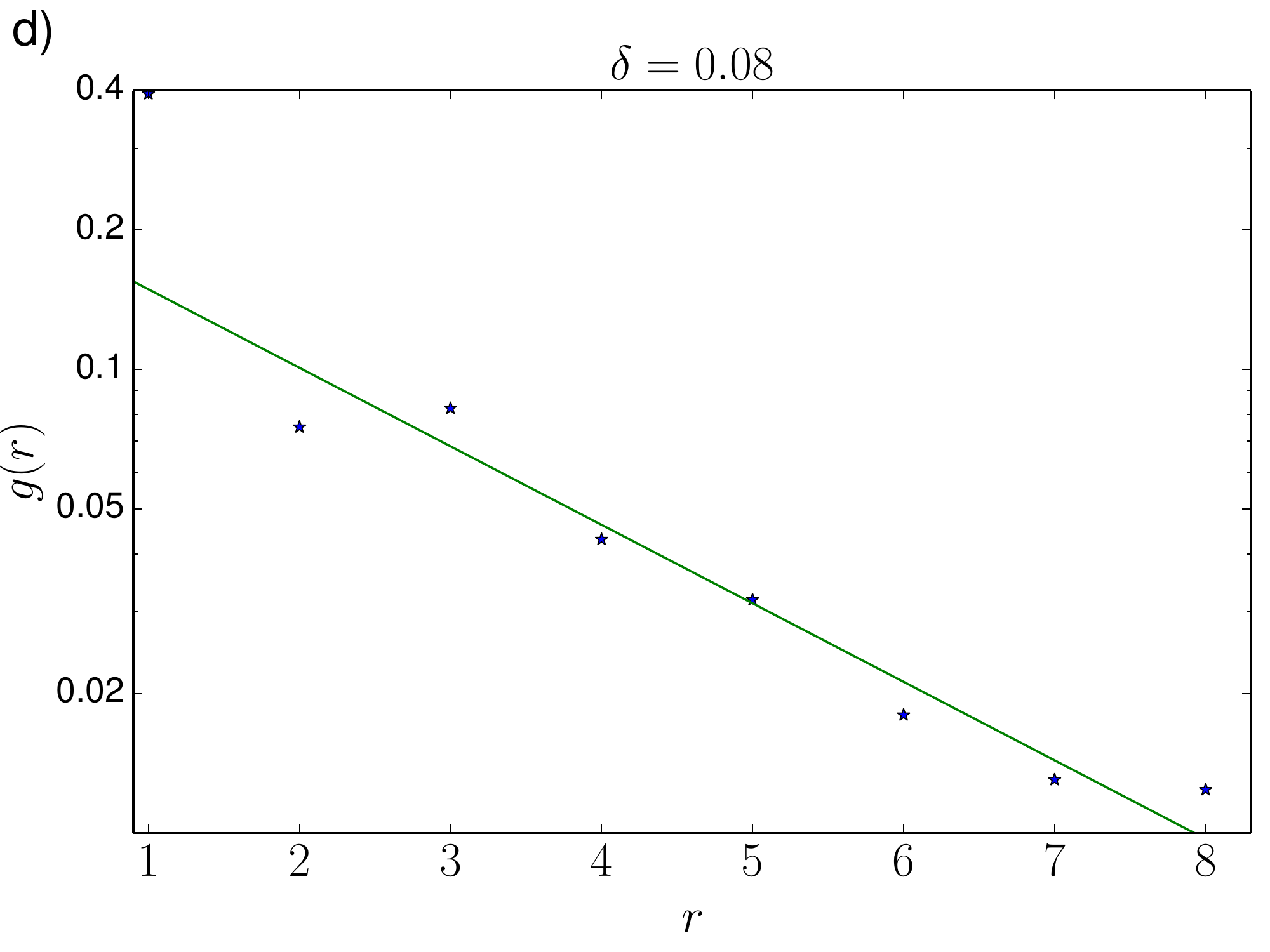}
\caption{Distance dependence of the spin--spin correlation function $g(r)$ for different hole contrentations $\delta=0.02$ (a), 0.04 (b), 0.06 (c) and 0.08 (d). A logarithmic scale is used on the vertical axis. The solid lines show fits to the Monte Carlo results. The following parameters have been assumed: $J=0.2t$, $kT=0.1t$, $t'=-0.27t$ and $t''=0.2t$.}%
\label{fig_corr_fun}
\end{center}
\end{figure}

Figure \ref{fig_corr_fun} shows the spin--spin correlation function $g(r)$ defined by Eq. \ref{corfun} for different hole concentrations. In order to describe the decay of the correlations we use a logarithmic scale on the vertical axis.
One can see in this figure that the character of this correlation changes very rapidly with the increase of the number of holes. For a very small concentration $\delta=0.02$ (Fig. \ref{fig_corr_fun}a) the correlation drops at a very short distance but then it is almost constant for larger distances, what indicates the presence of the long range antiferromagnetic order\footnote{The presented results were calculated for a 20$\times$20 system with periodic boundary conditions and therefore we cannot say anything about the behaviour of $g(r)$ at a very large distance.}. For a slightly higher doping $\delta=0.04$ (Fig. \ref{fig_corr_fun}b) we can observe an exponential decay at small distance, but then it slows down at larger distance changing into an algebraic decay. It suggests the presence of the quasi--long--range order. Finally, when we further increase the hole concentration to $\delta=0.06$ and $\delta=0.08$ (Figs. \ref{fig_corr_fun}c and \ref{fig_corr_fun}d, respectively), we observe an exponential decay at all distances, what means that the long range antiferromagnetic has been destroyed. It may suggest that the critical hole 
concentration in the Ising $t$--$J$ model is below 6\%, what is in a perfect agreement with experiments. This result, however, has been obtained on a relatively small cluster and should be confirmed by the finite--size scaling.

\begin{figure}
\begin{center}
\includegraphics[width=0.5\linewidth,natwidth=1030,natheight=773]{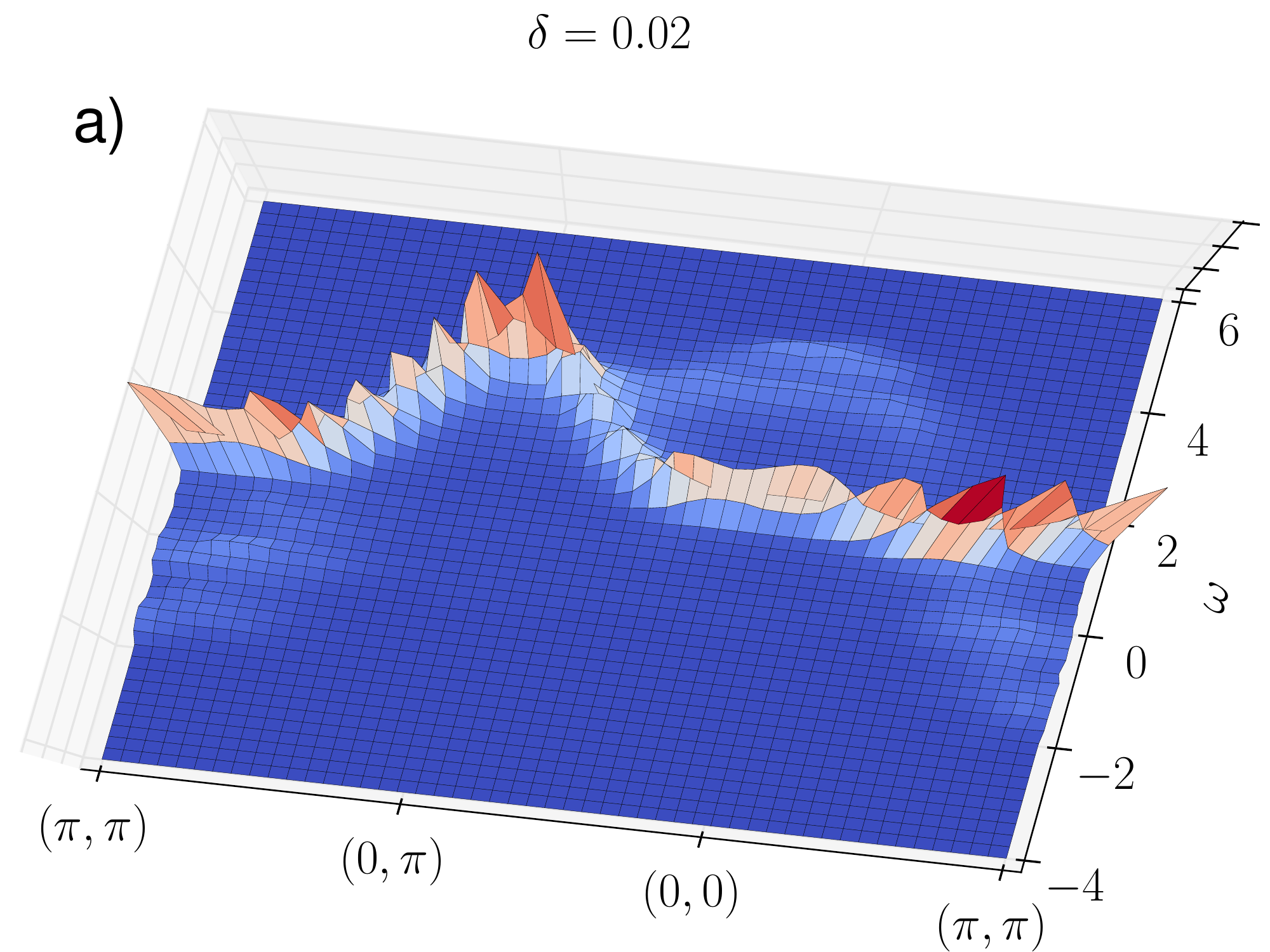}\includegraphics[width=0.5\linewidth,natwidth=1030,natheight=773]{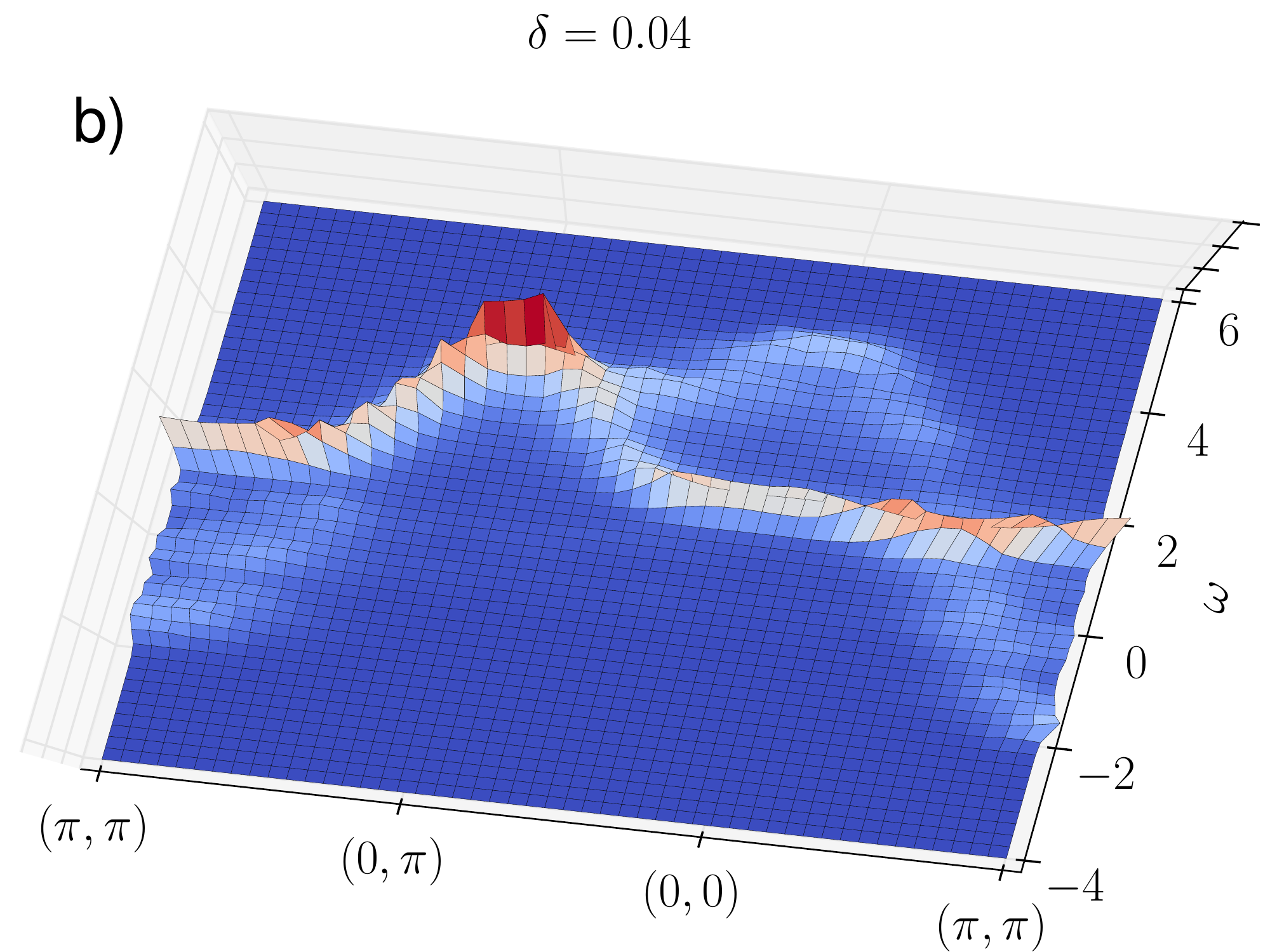}
\includegraphics[width=0.5\linewidth,natwidth=1030,natheight=773]{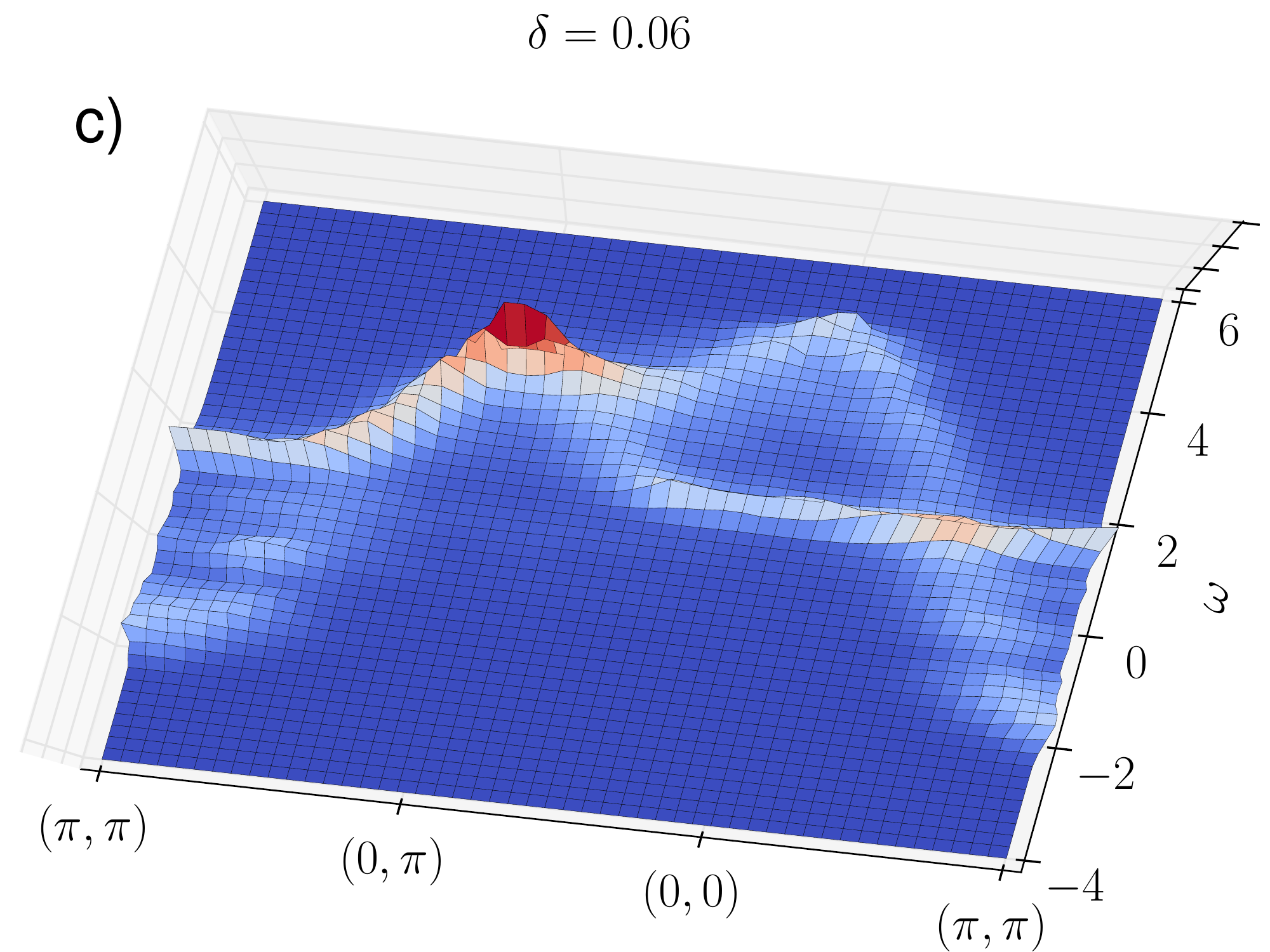}\includegraphics[width=0.5\linewidth,natwidth=1030,natheight=773]{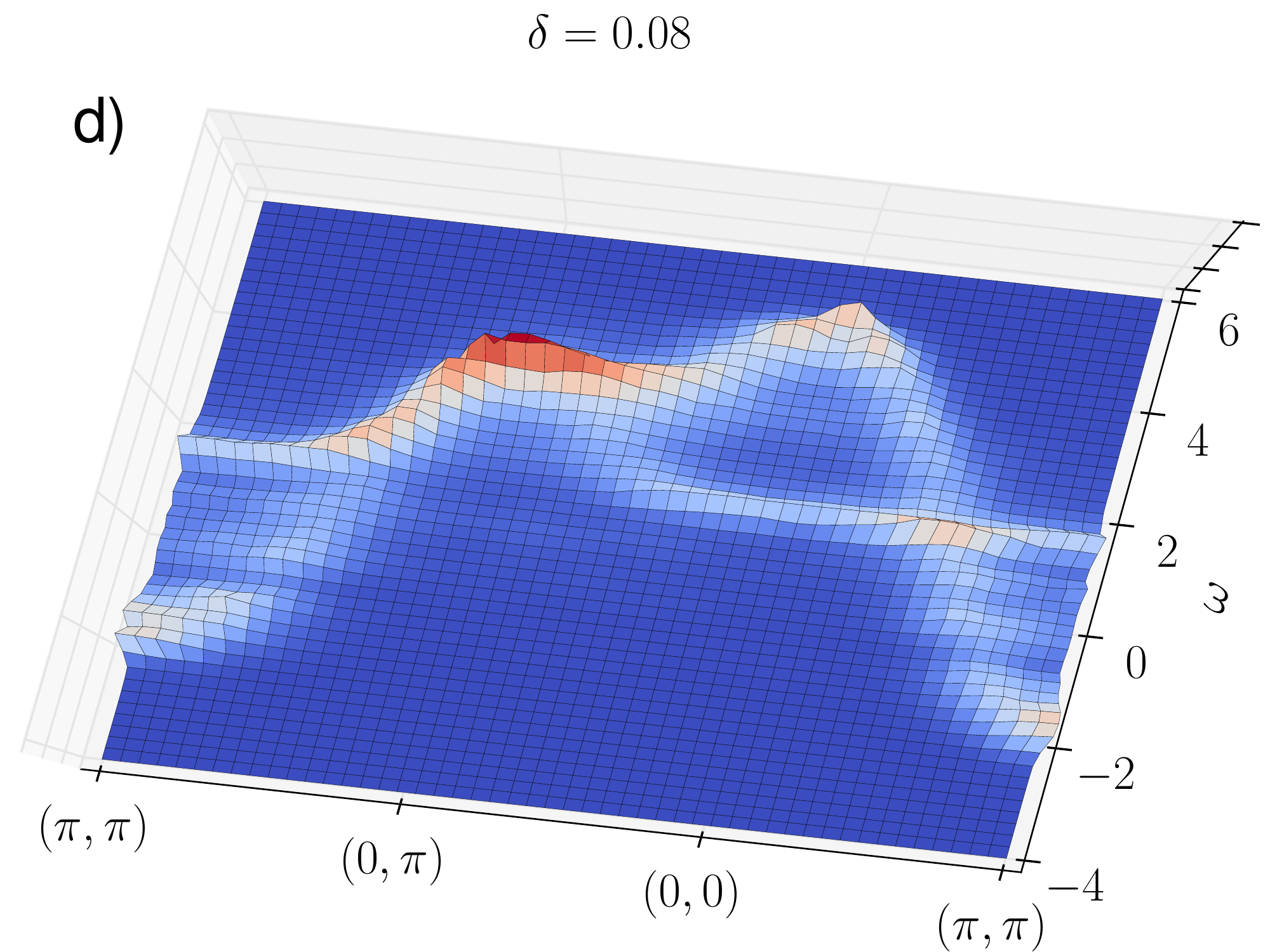}
\includegraphics[width=0.5\linewidth,natwidth=1030,natheight=773]{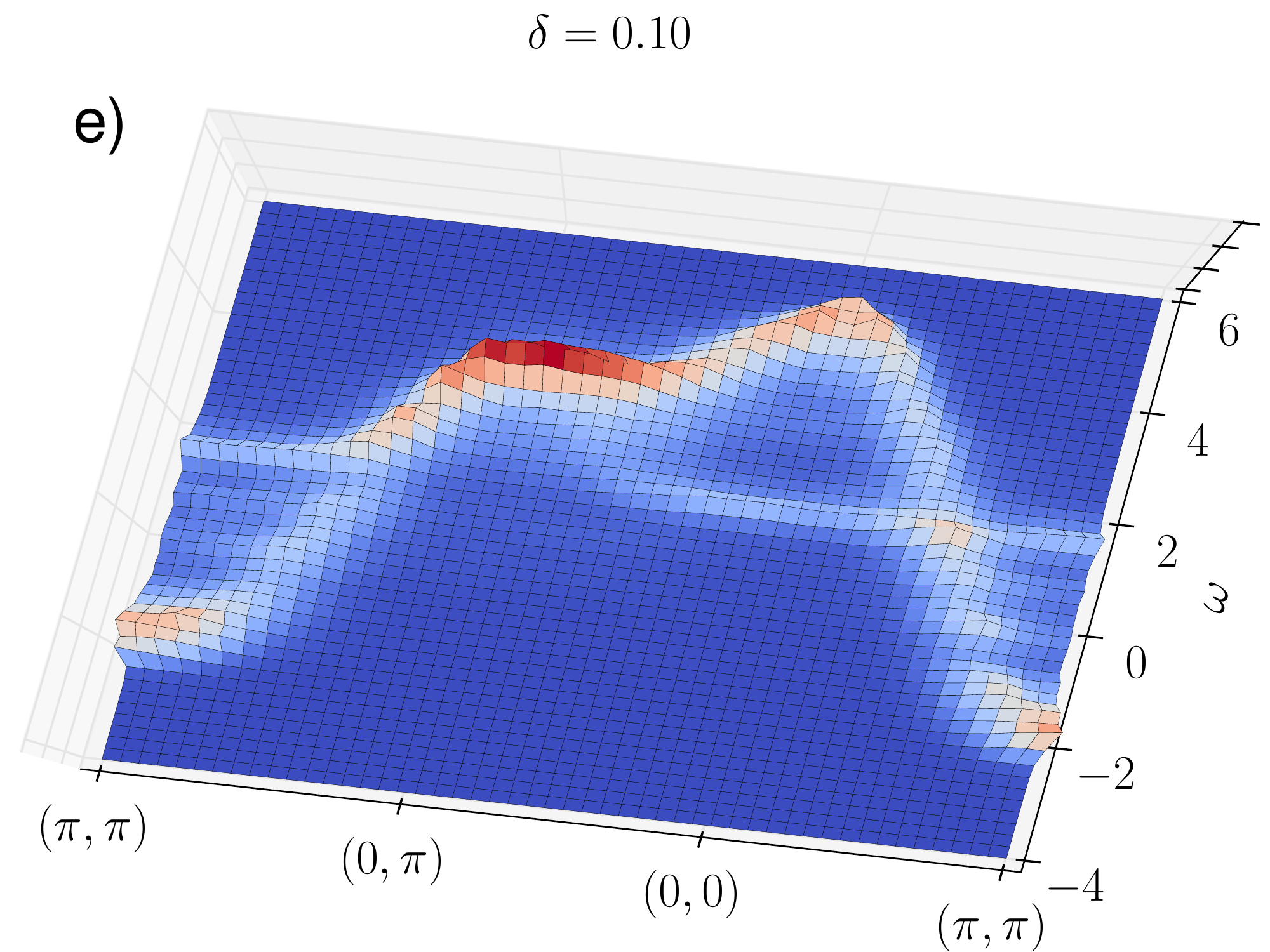}\includegraphics[width=0.5\linewidth,natwidth=1030,natheight=773]{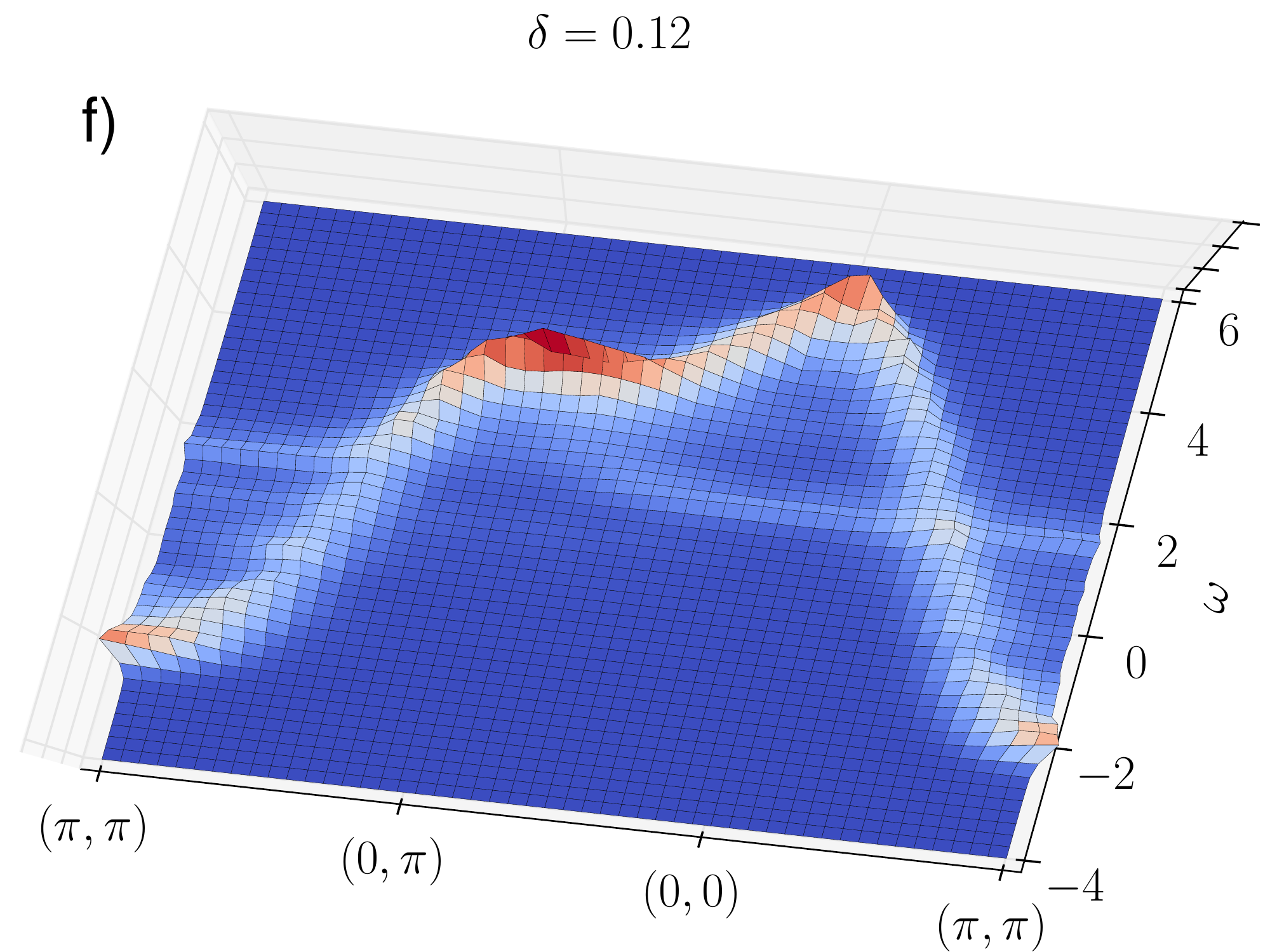}
\caption{Spectral functions for different hole concentration.}%
\label{spectr_funs}
\end{center}
\end{figure}

There is still, however, the question about the nature of the suppression of the long range 
antiferromagnetic order. Writing explicitly the $\lambda$--dependent term in Eqs. (\ref{tJ_Is}) and (\ref{hop_mat}) 
\begin{equation}
\lambda\sum_i\left[\left(\frac{1}{2}+Q^z_i\right)d^\dagger_{i\uparrow}d_{i\uparrow}+
\left(\frac{1}{2}-Q^z_i\right)d^\dagger_{i\downarrow}d_{i\downarrow}\right],
\end{equation}
one can see that in a perfect Neel state single holes which hop only to nearest neighbours 
are fully localized. The holes can gain kinetic energy by destruction of the 
antiferromagnetic order and forming a ferromagnetic region (spin polaron), but this
mechanism is effective only for a very small value of the exchange $J$ \cite{PhysRevB.85.245113,PhysRev.147.392}. Nonzero values of $t'$ and $t''$ allow holes to propagate
and gain some energy by intrasublattice hoppings. The same situation persists for a small
but finite concentration of holes. The hole spectral function calculated according to Eq. 
(\ref{adef}) for the hole concentration $\delta=0.02$ is presented in Fig. 
\ref{spectr_funs}a. It is exactly the spectral function for free electrons with the 
dispersion relation given by hoppings only to the second and third neighbours (i.e., within 
the same sublattice) with the hopping integrals $t'$ and $t''$, respectively.
When the hole concentration increases the potential gain from their mobility enhancement
would increase as well. And at some point it starts to be energetically favourable to
locally destroy the antiferromagnetic order and allow the holes to hop also to nearest 
neighbour sites. The evolution of this process can be observed in Figs. 
\ref{spectr_funs}b-f, where the contribution from electrons with the dispersion relation 
characteristic for nearest neighbour hopping becomes more and more pronounced. 
This results suggest that the main mechanism that is responsible for the suppression of
the long--range antiferromagnetic order is the competition between the hole mobility and
tendency towards minimization of the spin--spin exchange energy. This strong competition 
results directly from the no double occupancy constraint. This constraint, in connection with 
the lack of the spin--flip term in the Ising $t$--$J$ model prevents holes from hopping 
to nearest neighbour sites. If the gain is comparable to the antiferromagnetic ordering 
energy the total energy can be lowered by suppressing the order.

\section{Nagaoka polaron}
Since the proposed numerical approach to the Ising $t$--$J$ model does not require 
translational invariance it is well suited to study inhomogeneous systems. This feature 
allows to study the small--$J$ and small hole concentration limits of the $t$--$J$ model.
In the small--$J$ limit the dynamics of spins is much slower then 
the dynamics of charge carriers what may justify the approximation that leads to
the Ising version of the $t$--$J$ model, e.g., neglecting of the transverse spin 
components [Eq. (\ref{trans_spin})].

According to the Nagaoka theorem \cite{Nagaoka1965409, PhysRev.147.392} the infinite--$U$
Hubbard model with only a single hole has a 
fully spin--polarized ferromagnetic ground state. This regime corresponds to the $t$--$J$ 
model in the $J\rightarrow 0$ limit. The Nagaoka theorem is exceptional in the sense that 
it is one of very few rigorous results for strongly correlated systems. Unfortunately, 
its validity is limited only to a one particular case. The situation for large but finite
$U$ (what is equivalent to $J\gtrsim 0$) and/or small but finite hole concentration 
is much less clear. For finite $J$ there is the antiferromagnetic exchange energy
that competes with the kinetic energy of holes in a ferromagnetic spin background.
Therefore a single hole in a system with finite $J$ may lead to formation of a ferromagnetic
"bubble" that allows the hole to gain the kinetic energy. The rest of the system would have
anitferromagnetically ordered spins to minimize the exchange energy. The size of the 
ferromagnetic region would be determined by the competition between the exchange and kinetic 
energies in a way that guarantees a global minimum of the total \cite{PhysRevB.64.024411}. 

\begin{figure}
\begin{center}
\includegraphics[width=0.85\linewidth,natwidth=864,natheight=324]{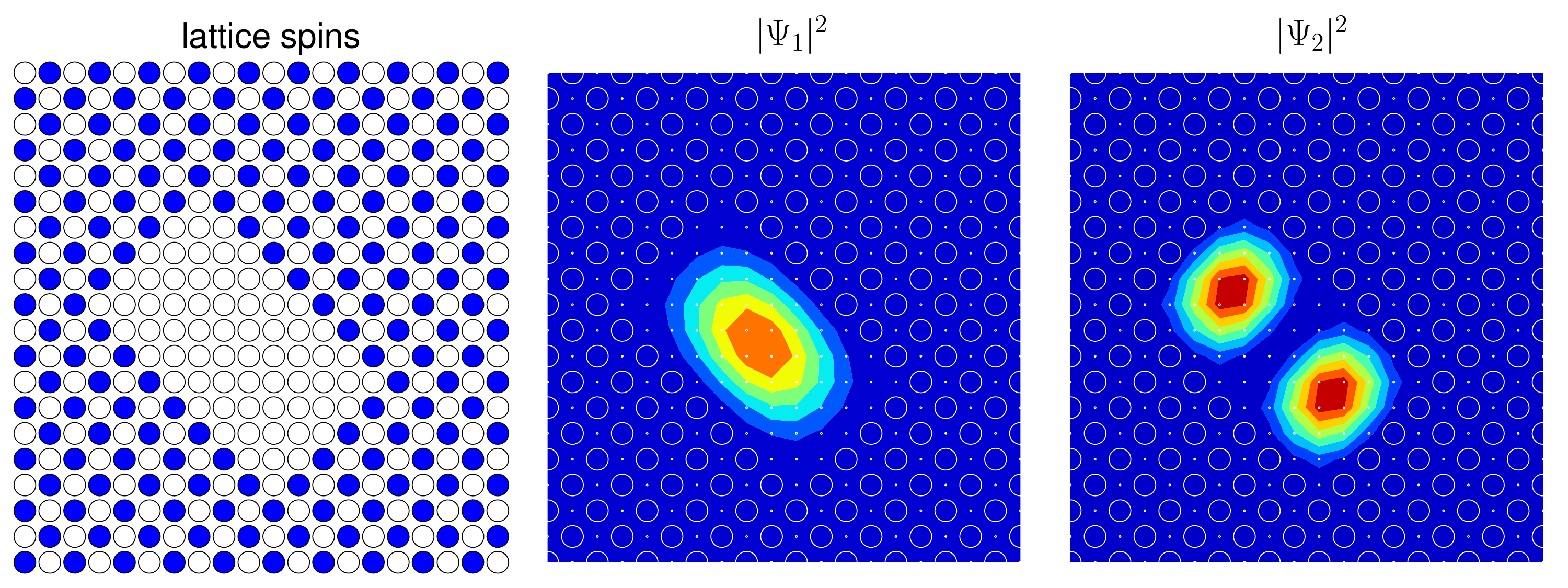}
\caption{Snapshot of the lattice spin configurations and false color plots of corresponding 
two lowest hole wave functions for $J=0.01$. In the leftmost picture filled (blue) circles
represent lattice spins pointing up and the empty (white) ones lattice spins pointing 
down. The white rectangular represent a ferromagnetic spin polaron.}
\label{bipolaron}
\end{center}
\end{figure}

The situation becomes more complicated with the increase of the number of holes.
Already for two holes results are ambiguous. In the small--$J$ regime the
question is whether the two holes will form a single bipolaron or two sparate polarons. Or
more generally, whether the holes form a bound state. The problem occurs because the
characteristic length scale in the small--$J$ regime, connected with the size of the 
ferromagnetic region, is large, beyond the limits of applicability of most of the fully 
quantum--mechanical approaches. One of the few methods which are capable to calculate 
properties of two holes in an antiferrormagnetic background is an accurate exact
diagonalization method, defined over a limited functional space (EDLFS) recently proposed
by Bon\v{c}a {\em et al}. in Refs. \cite{PhysRevB.76.035121,PhysRevLett.103.186401}.
The construction of the limited space starts from a Neel state with two holes located on 
neighboring lattice sites. Next, the kinetic and the spin--flip parts of the $t$--$J$ 
Hamiltonian are applied to generate the basis vectors. The ground state is then calculated
within the generated functional space by means of the Lancz\"os method. This approach allows
to study much larger systems than the standard exact diagonalization methods. Nevertheless, the maximum 
distance between the holes is limited by the size of the generated functional space. The 
problem is that the size of the ferromagnetic region which may contain the two holes
diverges with decreasing $J$ and at some point even this method does not allow to study 
the bipolaron problem. The value of $J$ below which the EDLFS method cannot give reliable 
results is about 0.04. On the other hand, the approximations that lead to the Ising $t$--$J$
model can be applied for arbitrarily small $J$ with its accuracy increasing with decreasing 
$J$. Therefore, we used a comparison of the 
results for the Ising $t$--$J$ model and the results of the EDLFS method applied to the 
full $t$--$J$ model to examine the validity of neglecting the transverse spin components
and to get insight into the physical meaning of this approximation. One important parameter
that can be compared is the average distance between two holes. In order to calculate 
its value in the Ising $t$--$J$ model we run Monte Carlo simulations for a system with two
holes. The difference between the present simulations and those carried out for the study
of the destruction of the antiferomagnetic order with the increase of the hole concentration
is that now we do not keep zero total magnetization. In the previous simulations each Monte
Carlo attempt consisted of two spin flips of two opposite lattice spins. Here the have two 
types of attempts: a transfer of a randomly chosen lattice spin from one site to another or 
a single spin flip. The latter kind of attempts does not conserve the total lattice 
magnetization. Figure \ref{bipolaron} shows a typical low--temperature configuration of the
lattice spins and two lowest corresponding hole wave functions $\Psi_1$ and $\Psi_2$. 
The average distance between the holes is calculated as
\begin{equation}
D=\left\langle \sum_{{\bm r}_1}\sum_{{\bm r}_2}|{\bm r}_1-{\bm r}_2|\:
{\cal P}({\bm r}_1,{\bm r}_2)\right\rangle,
\end{equation}
where ${\bm r}_1,\ {\bm r}_2$ run over all lattice sites,
\begin{equation}
{\cal P}({\bm r}_1,{\bm r}_2)=\left| \begin{array}{cc}
\Psi_1({\bm r}_1) & \Psi_1({\bm r}_2) \\
\Psi_2({\bm r}_1) & \Psi_2({\bm r}_2)
\end{array} \right|^2
\end{equation}
and $\langle\ldots\rangle$ denotes an average over the lattice spin configurations 
generated in a Monte Carlo run. It turned out that the distance between two holes in the
Ising $t$--$J$ model has the same dependence on $J$ as in the full $t$--$J$ model, but
the numeric prefactor is almost 30\% smaller. Namely, the distance $D(J)$ in the Ising 
$t$--$J$ model is given by $1.4\:r^{-0.27}$ and $1.97\:r^{-0.27}$ in the full $t$--$J$ 
model. The letter function has been obtained from a finite size scaling of the results of
the EDLFS method \cite{PhysRevB.85.245113}. The Monte Carlo simulations were carried out
for $J\leq 0.04$ and the EDLFS method was used for $J\geq 0.04$. For such a small value
of $J$ the Monte Carlo results indicated that the energy of one bipolaron is smaller than
that of two polarons, what suggests binding of the holes. The difference between the 
hole--hole distance in the full $t$--$J$ model and in its Ising version can be explained
by the approximations used in the Ising $t$--$J$ model: On the one hand, when the transverse
components of the spin operators are neglected the boundary between the ferromagnetic
"bubble" and its antiferromagnetic surroundings is impenetrable. On the other hand, the
spin flip term in the full $t$--$J$ model allows a hole to enter the antiferromagnetic
region. The movement of a hole through this region is accompanied by formation of a string
of defects in the antiferromagnetic order what strongly limits the range this penetration. 
The behaviour of the holes in these two models can be explained with the help of an analogy
to particles in quantum wells: the case of the Ising $t$--$J$ model it would be an infinite
rectangular quantum well, whereas in the full $t$--$J$ model the walls of the well would be
inclined outward, what would lead to a slightly broader wave function.
\subsection{Finite density of holes}
\begin{figure}
\begin{center}
\includegraphics[width=\linewidth,natwidth=1152,natheight=720]{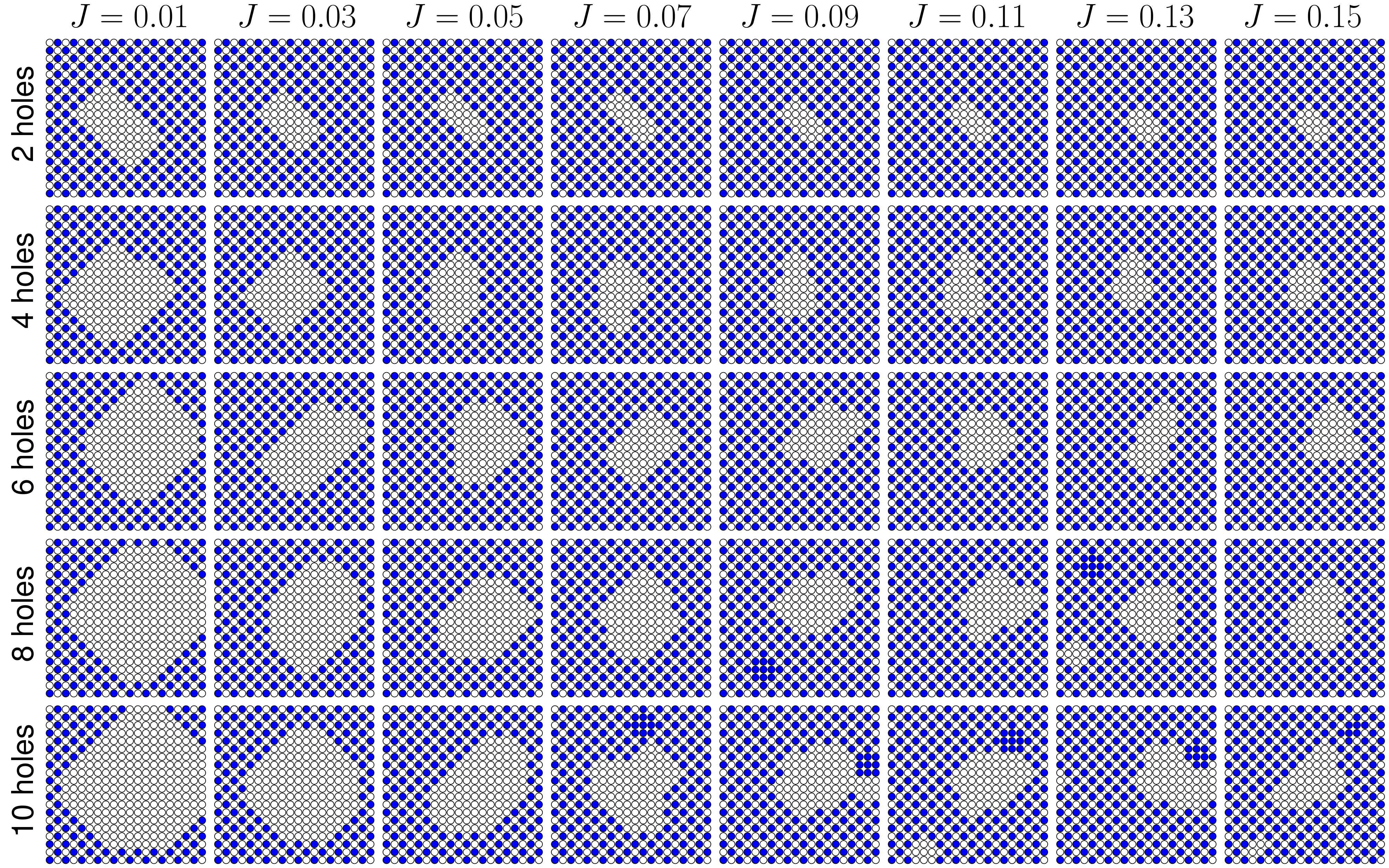}
\caption{Snapshots of the lattice spin configurations for different numbers of holes and
for different exchange interaction $J$. The meaning of the small filled and empty circles 
is the same as in Fig. \ref{bipolaron}. The main polaron is formed by lattice spins 
pointing down, whereas the dark areas in configuration for $J\ge 0.07$ and 8 and 10 holes
represent separate spin--up ferromagnetic polarons.}
\label{snapshots}
\end{center}
\end{figure}

\begin{figure}
\begin{center}
\includegraphics[width=0.75\linewidth,natwidth=576,natheight=432]{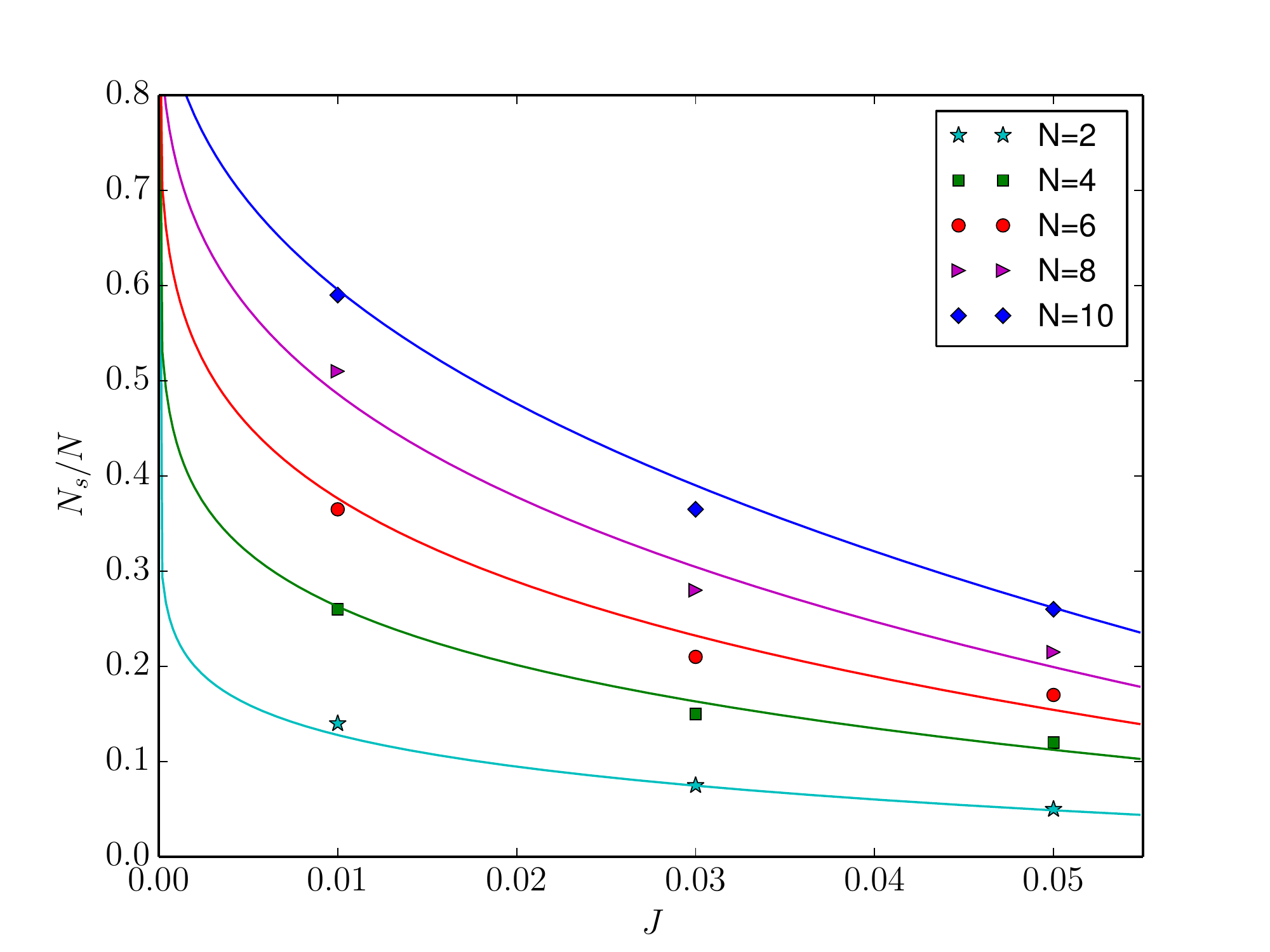}
\caption{Fraction of the total number of the lattice sites occupied by the ferromagnetic
polaron as a function of $J$ for different holes numbers.}
\label{polsize}
\end{center}
\end{figure}

For two holes in the small--$J$ limit it is energetically favourable to form a single 
ferromagnetic spin where the holes can move freely. Then, the question is whether this
scenario will hold for higher number of holes. Figure \ref{snapshots} show snapshots
of Monte Carlo simulations for up to 10 holes in a 20$\times$20 system for $J$ from 0.01
to 0.15. 
One can see there that for $J\ge 0.07$ and 8 and 10 holes multiple 
polarons are formed. This is, however, the region where the spin--flip processes may 
play a more important role and the validity of the approach may be questionable. Therefore,
we restrict ourselves to $J\leq 0.05$, similarly to the case of two holes. The Monte
Carlo studies in this regime show that the energy $E$ as a function of the number of holes
can be well fitted by $aN+b\sqrt{N}$, where $b$ is positive. It means that the function
$E(N)$ is concave and for all the studied hole concentrations ($N\le 10$) it is 
energetically
favourable to phase separate the system in a hole--rich ferromagnetic region and an
antiferromagnetic region without holes. As can be seen in Fig. \ref{snapshots}, the size
of the ferromagnetic "bubble" decreses with increasing $J$ what can be explained by the
increasing cost of broken antiferromagnetic bonds within the spin polaron. Fig. 
\ref{polsize} shows the size of the ferromagnetic polaron as a function of $J$.
The points from Monte Carlo simulations are there fitted by a function $N_s(J)/N=aJ^b$, 
where $N_s$ is the number number of the lattice sites occupied by the ferromagnetic
polaron, $N$ is the total number of the lattice sites and $a$ and $b$ are fitting 
parameters. For a fixed $J$ the size of the polaron increases with increasing hole 
concentration and at some point it includes all the lattice sites. This situation 
resembles the Nagaoka state, but for a finite number of holes. With decreasing $J$
the critical hole concentration decreases and in the $J\rightarrow 0$ limit 
($U\rightarrow \infty$) this state is converted into the standard Nagaoka state with 
vanishing hole concentration. The dependence of the critical hole concentration
as a function of $J$ can be well fitted by $\delta_t=0.44\:J^{0.53}$ what is very close to 
$\sqrt{J/2\pi}$, where the latter form can be easily derived by comparing the
kinetic energy of a few holes in an otherwise empty band and the exchange energy of the
lattice spins. If the hole concentration exceeds $\delta_t$ all the lattice spins are fully
polarized. For a hole concentration lower than $\delta_t$ the system is phase separated
into a hole--rich ferromagnetic part and a hole--depleted antiferromagnetic part. In this 
regime the size of the hole--rich ferromagnetic polaron (the number of lattice sites)
linearly depends on the hole concentration. The above results hold true in the limit 
of small $J$. As can be seen in Fig. \ref{snapshots} for $J\ge 0.07$ holes are confined 
to several separate polarons. In this regime, however, the validity of neglecting the
transverse spin components is questionable and the multi--polaron picture may not be
relevant to the isotropic $t$--$J$ model.

\section{Summary}
We have presented a representation of the of the $t$--$J$ model where the system is
described in terms of fermions interacting with static localized spins. Although it is a
slave--particle approach, in contrast with many similar approaches, the local no--double--occupancy constraint is rigorously taken into account. Within the proposed approach we have
shown that the long range antiferromagnetic order disappears already at the doping 
of the order of a few percent, what is in an agreement with the experimental data for 
high--$T_c$ superconductors. Additionally, we have demonstrated that the no--double--occupancy constraint is responsible for the destruction of the order. Since it is difficult 
to take this constraint into account in most of the analytical and numerical approaches
to the $t$--$J$ model, it explain why theoretical estimations of the critical dopant 
concentration usually give a significantly larger value. 

The proposed approach to the $t$--$J$ model does not require translational invariance of 
the system, what allow to use it to study inhomogeneous systems. Exploiting this feature
we have studied also formation of the Nagaoka polaron in the small--$J$ limit. The main 
difficulty in analysing the $t$--$J$ model in this limit is that a large size of the 
lattice is required to correctly describe the dynamics of holes. In the proposed approach,
however, the lattice spins are treated as classical variables, what allows to study systems
much larger than in fully quantum approaches like the Lancz\"os method, quantum Monte Carlo, 
DMRG, EDLFS, etc. Therefore, we were able to show that it is energetically favourable for
the system to segregate into the ferromagnetic hole-rich phase and hole-depleted 
antiferromagnetic phase. The size (surface) of the ferromagnetic bubble depends linearly
on the number of holes, while its dependence on $J$ is given by the square-root function.

\subsection*{Acknowledgement}

M.M.M. acknowledges support by the Polish National Science Center (NCN) under grant DEC-2013/11/B/ST3/00824. 
M.M. acknowledges support by the Polish National Science Center (NCN) under grant DEC-2013/09/B/ST3/01659. This work was carried out within the scope of the Bogoliubov--Infeld Programme.


\bibliographystyle{tPHM}
\bibliography{maska}

\begin{thebibliography}{10}
\newcommand{\noopsort}[1]{}
\newcommand{\printfirst}[2]{#1}
\newcommand{\singleletter}[1]{#1}
\newcommand{\switchargs}[2]{#2#1}
\providecommand{\url}[1]{\normalfont{#1}}
\providecommand{\urlprefix}{Available at }

\bibitem{anderson1959}
P.W. Anderson, Phys. Rev. 115 (1959), pp. 2--13,
  \urlprefix\url{http://link.aps.org/doi/10.1103/PhysRev.115.2}.

\bibitem{hubbard}
J. Hubbard, Proceedings of the Royal Society of London. Series A. Mathematical
  and Physical Sciences 276 (1963), pp. 238--257,
  \urlprefix\url{http://rspa.royalsocietypublishing.org/content/276/1365/238.abstract}.

\bibitem{spalek1977}
K.A. Chao, J. Spalek, and A.M. Oles, Journal of Physics C: Solid State Physics
  10 (1977), p. L271,
  \urlprefix\url{http://stacks.iop.org/0022-3719/10/i=10/a=002}.

\bibitem{spalek1978}
K.A. Chao, J. Spa\l{}ek, and A.M. Ole\ifmmode~\acute{s}\else  \'{s}\fi{}, Phys.
  Rev. B 18 (1978), pp. 3453--3464,
  \urlprefix\url{http://link.aps.org/doi/10.1103/PhysRevB.18.3453}.

\bibitem{Baskaran1987973}
G. Baskaran, Z. Zou, and P. Anderson, Solid State Communications 63 (1987), pp.
  973 -- 976,
  \urlprefix\url{http://www.sciencedirect.com/science/article/pii/0038109887906429}.

\bibitem{PhysRevLett.58.2790}
P.W. Anderson, G. Baskaran, Z. Zou, and T. Hsu, Phys. Rev. Lett. 58 (1987), pp.
  2790--2793,
  \urlprefix\url{http://link.aps.org/doi/10.1103/PhysRevLett.58.2790}.

\bibitem{Kotliar88}
G. Kotliar and J. Liu, Phys. Rev. B 38 (1988), pp. 5142--5145,
  \urlprefix\url{http://link.aps.org/doi/10.1103/PhysRevB.38.5142}.

\bibitem{JPSJ.57.2768}
Y. Suzumura, Y. Hasegawa, and H. Fukuyama, Journal of the Physical Society of
  Japan 57 (1988), pp. 2768--2778,
  \urlprefix\url{http://jpsj.ipap.jp/link?JPSJ/57/2768/}.

\bibitem{RevModPhys.78.17}
P.A. Lee, N. Nagaosa, and X.G. Wen, Rev. Mod. Phys. 78 (2006), pp. 17--85,
  \urlprefix\url{http://link.aps.org/doi/10.1103/RevModPhys.78.17}.

\bibitem{mmfk2009}
M.M. Ma\'ska, M. Mierzejewski, A. Ferraz, and E.A. Kochetov, Journal of
  Physics: Condensed Matter 21 (2009), p. 045703,
  \urlprefix\url{http://stacks.iop.org/0953-8984/21/i=4/a=045703}.

\bibitem{ribeiro2005}
T.C. Ribeiro and X.G. Wen, Phys. Rev. Lett. 95 (2005), p. 057001,
  \urlprefix\url{http://link.aps.org/doi/10.1103/PhysRevLett.95.057001}.

\bibitem{fku2007}
A. Ferraz, E. Kochetov, and B. Uchoa, Phys. Rev. Lett. 98 (2007), p. 069701,
  \urlprefix\url{http://link.aps.org/doi/10.1103/PhysRevLett.98.069701}.

\bibitem{mm1}
M.M. Ma\ifmmode~\acute{s}\else  \'{s}\fi{}ka and K. Czajka, Phys. Rev. B 74
  (2006), p. 035109,
  \urlprefix\url{http://link.aps.org/doi/10.1103/PhysRevB.74.035109}.

\bibitem{mm2}
M.M. Ma\'ska and K. Czajka, physica status solidi (b) 242 (2005), pp. 479--483,
  \urlprefix\url{http://dx.doi.org/10.1002/pssb.200460067}.

\bibitem{mm3}
M.M. Ma\ifmmode~\acute{s}\else  \'{s}\fi{}ka, R. Lema\ifmmode~\acute{n}\else
  \'{n}\fi{}ski, J.K. Freericks, and C.J. Williams, Phys. Rev. Lett. 101
  (2008), p. 060404,
  \urlprefix\url{http://link.aps.org/doi/10.1103/PhysRevLett.101.060404}.

\bibitem{PhysRevB.52.4597}
J. Ba{\l}a, A.M. Ole\ifmmode~\acute{s}\else  \'{s}\fi{}, and J. Zaanen, Phys.
  Rev. B 52 (1995), pp. 4597--4606,
  \urlprefix\url{http://link.aps.org/doi/10.1103/PhysRevB.52.4597}.

\bibitem{PhysRevB.54.R12653}
T. Xiang and J.M. Wheatley, Phys. Rev. B 54 (1996), pp. R12653--R12656,
  \urlprefix\url{http://link.aps.org/doi/10.1103/PhysRevB.54.R12653}.

\bibitem{PhysRevB.54.10125}
B. Kyung and R.A. Ferrell, Phys. Rev. B 54 (1996), pp. 10125--10130,
  \urlprefix\url{http://link.aps.org/doi/10.1103/PhysRevB.54.10125}.

\bibitem{PhysRevLett.90.067001}
T.K. Lee, C.M. Ho, and N. Nagaosa, Phys. Rev. Lett. 90 (2003), p. 067001,
  \urlprefix\url{http://link.aps.org/doi/10.1103/PhysRevLett.90.067001}.

\bibitem{metropolis}
N. Metropolis, A.W. Rosenbluth, M.N. Rosenbluth, A.H. Teller, and E. Teller,
  The Journal of Chemical Physics 21 (1953), pp. 1087--1092,
  \urlprefix\url{http://scitation.aip.org/content/aip/journal/jcp/21/6/10.1063/1.1699114}.

\bibitem{PhysRevLett.60.2793}
S. Schmitt-Rink, C.M. Varma, and A.E. Ruckenstein, Phys. Rev. Lett. 60 (1988),
  pp. 2793--2796,
  \urlprefix\url{http://link.aps.org/doi/10.1103/PhysRevLett.60.2793}.

\bibitem{PhysRevLett.60.740}
B.I. Shraiman and E.D. Siggia, Phys. Rev. Lett. 60 (1988), pp. 740--743,
  \urlprefix\url{http://link.aps.org/doi/10.1103/PhysRevLett.60.740}.

\bibitem{PhysRevB.37.1597}
S.A. Trugman, Phys. Rev. B 37 (1988), pp. 1597--1603,
  \urlprefix\url{http://link.aps.org/doi/10.1103/PhysRevB.37.1597}.

\bibitem{PhysRevB.58.15160}
P. Wr\'obel and R. Eder, Phys. Rev. B 58 (1998), pp. 15160--15176,
  \urlprefix\url{http://link.aps.org/doi/10.1103/PhysRevB.58.15160}.

\bibitem{PhysRevB.41.9049}
E. Dagotto, R. Joynt, A. Moreo, S. Bacci, and E. Gagliano, Phys. Rev. B 41
  (1990), pp. 9049--9073,
  \urlprefix\url{http://link.aps.org/doi/10.1103/PhysRevB.41.9049}.

\bibitem{PhysRevB.42.10706}
P. Prelovek, I. Sega, and J. Bona, Phys. Rev. B 42 (1990), pp. 10706--10713,
  \urlprefix\url{http://link.aps.org/doi/10.1103/PhysRevB.42.10706}.

\bibitem{PhysRevB.44.317}
G. Martinez and P. Horsch, Phys. Rev. B 44 (1991), pp. 317--331,
  \urlprefix\url{http://link.aps.org/doi/10.1103/PhysRevB.44.317}.

\bibitem{PhysRevB.47.14267}
D. Poilblanc, T. Ziman, H.J. Schulz, and E. Dagotto, Phys. Rev. B 47 (1993),
  pp. 14267--14279,
  \urlprefix\url{http://link.aps.org/doi/10.1103/PhysRevB.47.14267}.

\bibitem{PhysRevB.52.R15711}
P.W. Leung and R.J. Gooding, Phys. Rev. B 52 (1995), pp. R15711--R15714,
  \urlprefix\url{http://link.aps.org/doi/10.1103/PhysRevB.52.R15711}.

\bibitem{PhysRevB.62.15480}
M. Brunner, F.F. Assaad, and A. Muramatsu, Phys. Rev. B 62 (2000), pp.
  15480--15492,
  \urlprefix\url{http://link.aps.org/doi/10.1103/PhysRevB.62.15480}.

\bibitem{PhysRevB.40.2610}
C. Jayaprakash, H.R. Krishnamurthy, and S. Sarker, Phys. Rev. B 40 (1989), pp.
  2610--2613, \urlprefix\url{http://link.aps.org/doi/10.1103/PhysRevB.40.2610}.

\bibitem{0953-8984-6-27-022}
Y.A. Izyumov, B.M. Letfulov, and E.V. Shipitsyn, Journal of Physics: Condensed
  Matter 6 (1994), p. 5137,
  \urlprefix\url{http://stacks.iop.org/0953-8984/6/i=27/a=022}.

\bibitem{PhysRevB.83.104512}
J. Jedrak and J. Spa\l{}ek, Phys. Rev. B 83 (2011), p. 104512,
  \urlprefix\url{http://link.aps.org/doi/10.1103/PhysRevB.83.104512}.

\bibitem{PhysRevB.79.214519}
J. Kaczmarczyk and J. Spa\l{}ek, Phys. Rev. B 79 (2009), p. 214519,
  \urlprefix\url{http://link.aps.org/doi/10.1103/PhysRevB.79.214519}.

\bibitem{PhysRevB.81.073108}
J. Jedrak and J. Spa\l{}ek, Phys. Rev. B 81 (2010), p. 073108,
  \urlprefix\url{http://link.aps.org/doi/10.1103/PhysRevB.81.073108}.

\bibitem{0295-5075-52-1-087}
C.H. Cheng and T.K. Ng, EPL (Europhysics Letters) 52 (2000), p.~87,
  \urlprefix\url{http://stacks.iop.org/0295-5075/52/i=1/a=087}.

\bibitem{PhysRevB.59.8943}
Z.Y. Weng, D.N. Sheng, and C.S. Ting, Phys. Rev. B 59 (1999), pp. 8943--8955,
  \urlprefix\url{http://link.aps.org/doi/10.1103/PhysRevB.59.8943}.

\bibitem{PhysRevB.44.12077}
T.I. Ivanov, Phys. Rev. B 44 (1991), pp. 12077--12079,
  \urlprefix\url{http://link.aps.org/doi/10.1103/PhysRevB.44.12077}.

\bibitem{PhysRevB.41.2653}
C.L. Kane, P.A. Lee, T.K. Ng, B. Chakraborty, and N. Read, Phys. Rev. B 41
  (1990), pp. 2653--2656,
  \urlprefix\url{http://link.aps.org/doi/10.1103/PhysRevB.41.2653}.

\bibitem{PhysRevLett.80.2393}
M. Fleck, A.I. Liechtenstein, A.M. Ole\ifmmode~\acute{s}\else  \'{s}\fi{}, L.
  Hedin, and V.I. Anisimov, Phys. Rev. Lett. 80 (1998), pp. 2393--2396,
  \urlprefix\url{http://link.aps.org/doi/10.1103/PhysRevLett.80.2393}.

\bibitem{PhysRevB.85.245113}
M.M. Ma\ifmmode~\acute{s}\else  \'{s}\fi{}ka, M. Mierzejewski, E.A. Kochetov,
  L. Vidmar, J. Bon\ifmmode~\check{c}\else  \v{c}\fi{}a, and O.P. Sushkov,
  Phys. Rev. B 85 (2012), p. 245113,
  \urlprefix\url{http://link.aps.org/doi/10.1103/PhysRevB.85.245113}.

\bibitem{PhysRevB.64.024411}
S.R. White and I. Affleck, Phys. Rev. B 64 (2001), p. 024411,
  \urlprefix\url{http://link.aps.org/doi/10.1103/PhysRevB.64.024411}.

\bibitem{PhysRevB.76.035121}
J. Bon\ifmmode~\check{c}\else  \v{c}\fi{}a, S. Maekawa, and T. Tohyama, Phys.
  Rev. B 76 (2007), p. 035121,
  \urlprefix\url{http://link.aps.org/doi/10.1103/PhysRevB.76.035121}.

\bibitem{PhysRevLett.103.186401}
L. Vidmar, J. Bon\ifmmode~\check{c}\else  \v{c}\fi{}a, S. Maekawa, and T.
  Tohyama, Phys. Rev. Lett. 103 (2009), p. 186401,
  \urlprefix\url{http://link.aps.org/doi/10.1103/PhysRevLett.103.186401}.

\bibitem{PhysRev.147.392}
Y. Nagaoka, Phys. Rev. 147 (1966), pp. 392--405,
  \urlprefix\url{http://link.aps.org/doi/10.1103/PhysRev.147.392}.

\bibitem{Nagaoka1965409}
Y. Nagaoka, Solid State Communications 3 (1965), pp. 409 -- 412,
  \urlprefix\url{http://www.sciencedirect.com/science/article/pii/0038109865902668}.

\end{thebibliography}

\end{document}